\documentclass[10pt,twocolumn,letterpaper]{article}

\usepackage{cvpr}              % To produce the CAMERA-READY version

\definecolor{cvprblue}{rgb}{0.21,0.49,0.74}
\usepackage{caption}
\usepackage[pagebackref,breaklinks,colorlinks,allcolors=cvprblue]{hyperref}
\usepackage{multirow}
\usepackage{makecell}
\usepackage{bbding}
\usepackage{mdframed}
\usepackage{tabulary}
\usepackage{colortbl}
\usepackage{tcolorbox}
\usepackage{utfsym}
\usepackage{tikz}
\usepackage{soul}
\usepackage[hang,flushmargin]{footmisc}

\def\paperID{15255} % *** Enter the Paper ID here
\def\confName{CVPR}
\def\confYear{2026}

\newcommand{\edited}[1]{{\textcolor{black}{#1}}}

\newcommand{\model}[1]{{OmniAbnorm-CT}}
\newcommand{\dataset}[1]{{OmniAbnorm-CT-14K}}

\definecolor{lightskyblue}{rgb}{0.53, 0.81, 0.98}
\definecolor{lightgray}{RGB}{242,242,242}
\definecolor{lightbrown}{RGB}{242,225,216}
\definecolor{lightblue}{RGB}{246,248,252}

\title{Rethinking Whole-Body CT Image Interpretation: \\ An Abnormality-Centric Approach}

\author{Ziheng Zhao$^{1, 2*}$, Lisong Dai$^{3*}$, Ya Zhang$^{1, 2}$, Weidi Xie$^{1, 2\dagger}$, Yanfeng Wang$^{1, 2\dagger}$ \\[4pt]
$^{1}$School of Artificial Intelligence, Shanghai Jiao Tong University \\[2pt]
$^{2}$Shanghai Artificial Intelligence Laboratory\\[2pt]
$^{3}$Department of Radiology, Renmin Hospital of Wuhan University\\
[6pt]
\url{https://github.com/zhaoziheng/OmniAbnorm-CT}
}

\begin{document}

\maketitle

% The followings are for CVPR

\begin{abstract}
Automated interpretation of CT images—particularly localizing and describing abnormal findings across multi-plane and whole-body scans—remains a significant challenge in clinical radiology. This work aims to address this challenge through four key contributions: 
(i) On taxonomy, we collaborate with senior radiologists to propose a comprehensive hierarchical classification system, with 404 representative abnormal findings across all body regions;
(ii) On data, we contribute a dataset containing over 14.5K CT images from multiple planes and all human body regions, and meticulously provide grounding annotations for over 19K abnormalities, each linked to the detailed description and cast into the taxonomy;
(iii) On model development, we propose \textbf{\model~}, which can automatically ground and describe abnormal findings on multi-plane and whole-body CT images based on text queries, while also allowing flexible interaction through visual prompts;
(iv) On evaluation, we establish three representative tasks based on real clinical scenarios, and introduce a clinically grounded metric to assess abnormality descriptions.
Through extensive experiments, we show that \model~ can significantly outperform existing methods in both internal and external validations, and across all the tasks.
\end{abstract}

\begin{figure*}[t]
  \centering
  \includegraphics[width=\textwidth]{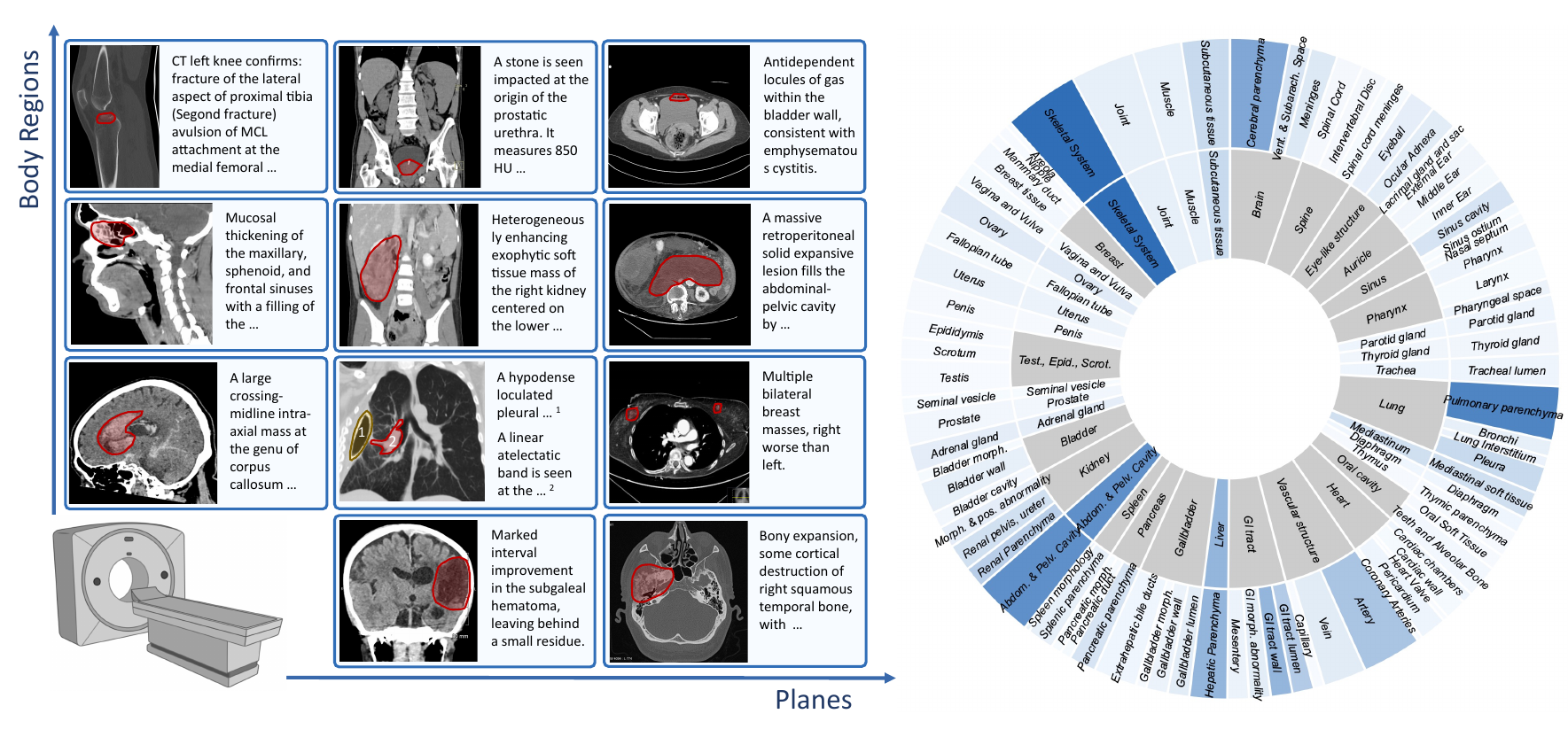} \\
  \vspace{-3pt}
  \captionof{figure}{
    This work introduces \textbf{\dataset~}, the first large-scale dataset for grounding and describing abnormal findings in CT images.
    \emph{\textbf{Left:}} \dataset~ contains 14.5K multi-plane, whole-body CT images, covering 349 representative abnormal findings across 82 anatomical structures and 40 major systems or organs.
    \emph{\textbf{Right:}} Distribution of the dataset across anatomical structures and major systems or organs, with darker blue indicating higher sample density.
    }
  \vspace{-.3cm}
 \label{fig:teaser}
\end{figure*}    
\section{Introduction}

Computed Tomography (CT) imaging has become a cornerstone of modern medicine, 
with over 450 million scans performed annually worldwide.
The interpretation of these scans, particularly the generation of radiology reports, plays a crucial role in clinical decision-making. 
However, systematically identifying and characterizing abnormal findings in whole-body CT images remains a cognitively demanding and time-consuming task for radiologists~\cite{rao2025multimodal}.

Advances in artificial intelligence (AI) have shown promising progress in CT image interpretation, primarily along two directions:
{\em first}, models developed for organ segmentation~\cite{Totalsegmentator,jaus2023towards} enable precise anatomical labeling and organ-grounded report generation~\cite{zhang2024radgenome,chen2025large}. 
Yet a critical gap remains: segmenting organs alone is insufficient for clinical utility. What ultimately matters to clinicians are abnormalities, lesions, and any anomalies that inform diagnosis and treatment;
{\em second}, models for automatic report generation have been trained on the recent datasets with paired CT scans and radiology reports, describing findings and impressions~\cite{chen2024bimcv,hamamci2024developing}. 
However, these datasets are typically limited to specific anatomical regions, {\em e.g.}, chest CT, and often lack explicit visual grounding, which reduces explainability and increases the risk of hallucinations. 
To address these challenges, we aim to develop a system that \textbf{detects, localizes, and describes all abnormal findings across multi-plane, whole-body CT images} automatically. This paradigm shift moves beyond regional reporting and anatomical segmentation, aligning AI interpretation more closely with practical radiological demands.

One crucial barrier to developing such a system is the absence of a unified, clinically meaningful taxonomy of abnormalities, which hinders the definition of task scope and continuous benchmarks. In this paper, we collaborated with 7 senior radiologists from 3 centers~(each with 10-16 years of experience) to construct a hierarchical taxonomy covering 404 representative abnormal findings, organized across 40 major anatomical regions and 82 sub-regions.

Based on the taxonomy, we introduce \textbf{\dataset}, the first large-scale dataset designed for abnormality grounding and description on multi-plane whole-body CT imaging. 
It comprises 14.5K CT images from Radiopedia~\cite{radiopaedia}, covering axial, coronal, and sagittal planes and diverse anatomical regions. 
All images and the paired reports are contributed and rigorously reviewed by worldwide radiologists. 
We invite 4 board-certified radiologists to manually annotate around 19K abnormal findings on the images in the format of either bounding box or segmentation masks. 
All regional annotations are linked to the corresponding report descriptions, and further cast into our proposed taxonomy. 
These annotations cover 349 out of the 404 representative abnormal findings (86\%) defined in our taxonomy across whole-body CT images. 
By providing high-quality, clinically relevant annotations at scale, 
this dataset addresses the critical data limitations in the field, 
caused by privacy constraints or high annotation cost.
All our annotations will be released to the community, while the images can be accessed for research purposes after application\footnote{\url{https://radiopaedia.org/licence\#data-sets}}.

% Leveraging \dataset~, we develop \textbf{\model~}, a novel system for grounded CT image interpretation. 
% By bridging a multi-modal language model with a specialized segmentation module in a bidirectional workflow, 
% it enables dynamic invocation of the segmentation module during the text generation, 
% and grounded abnormality interpretation based on the segmentation results.
% In contrast to prior approaches—which often lack grounding abilities, generalizability across planes and body regions, or flexibility in prompt handling—\model~ enables automatic localization and description of abnormalities across multi-plane, whole-body CT scans. 
% More importantly, \model~ is designed to robustly interpret diverse textual references and visual prompts, enabling interactive and clinically adaptive use cases aligned with radiologists' workflows.
% \xiaoman{I feel this paragraph is not very clear to me, not able to give me a concrete understanding of what the model actually does.}

Leveraging this dataset, we develop \textbf{\model~}, a novel system for grounded CT image interpretation. 
By bridging a vision language model (VLM) with the specialized segmentation module in a bidirectional workflow,
it enables the VLM to dynamiclly invoke the segmentation module during generation, 
and generate abnormality interpretation grounded on the segmentation results.
In contrast to prior approaches, which often lack grounding abilities, generalizability across planes and body regions, or flexibility in prompt handling, \model~ enables automatic localization and description of abnormalities across multi-plane, whole-body CT images. 
Moreover, \model~ is designed to robustly interpret diverse textual and visual prompts, enabling interactive and adaptive use cases aligned with radiologists' clinical workflows.

% For a comprehensive evaluation,
% we define three clinically relevant tasks: (i) grounded report generation, (ii) text-guided grounded report generation, and (iii) visual-prompted report generation. 
% Experiment results show that \model~ significantly outperforms existing baselines across all tasks, marking a substantial step toward explainable, abnormality-centric CT imaging interpretation
% \xiaoman{the tasks appear without any context, maybe add short definition/examples of each tasks?}

For a comprehensive evaluation,
we first define three clinically relevant tasks: 
(i) \textbf{grounded report generation} to detect, localize, and describe all findings;
(ii) \textbf{text-guided grounded report generation} to detect, localize, and describe abnormalities based on textual queries about specific clinical concerns, 
and (iii) \textbf{visual-prompted report generation} to describe a particular abnormality indicated by a visual prompt.
Then, we propose a clinically grounded metric for report evaluation, named \textbf{AbnormRubric}.
In contrast to existing metrics relying on simple matching~\cite{bleu,rouge,banerjee2005meteor} or overall semantic similarity~\cite{zhangbertscore,zhao2024ratescore,delbrouck2024radgraph}, 
AbnormRubric customizes the rubric for each abnormality category, ensure alignment with clinical judgement.
Implemented using a large language model (LLM) as a judge, 
it explicitly quantifies recall and precision in abnormality detection, along with the descriptive accuracy of abnormalitis across four key radiological attributions: location, morphology, density, and size.
Experiment results show that \model~ significantly outperforms existing baselines across all tasks, marking a substantial step toward explainable, abnormality-centric CT imaging interpretation.

In summary, our contributions to advance abnormality-centric grounded CT image interpretation are: 
(i) A comprehensive taxonomy for abnormalities on CT images, presented in Section~\ref{sec:taxonomy}; 
(ii) \textbf{\dataset~}, the first dataset for abnormality grounding and description across multi-plane whole-body CT images, detailed in Section~\ref{sec:data}; 
(iii) \textbf{\model~}, a system for automatically localizing and reporting abnormalities, illustrated in Section~\ref{sec:model}; 
(iv) A benchmark with three representative tasks and a clinically grounded metrics for report evaluation, 
demonstrating the superiority of \model~ in Section~\ref{sec:experiment}.

\section{Related Work}
\textbf{Medical image segmentation} aims to delineate clinically meaningful regions on medical images.
For radiology, most datasets focus on annotations for important organs and anatomical structures~\cite{Totalsegmentator,abdomenatlas,FLARE,AMOS22} or lesions~\cite{heller2021kidney,MSD,BraTS2023GLI,de2024uls23,yan2018deep}.
The past decade witnessed both remarkable success in specialist segmentation models~\cite{nnUNET,UNet,zhou2023nnformer,zhou2019unet++} and a recent paradigm shift towards building generalist models~\cite{MedSAM,zhao2024foundation,zhao2025large,du2024segvol}.
However, few works contribute to the grounding of abnormal image findings, a broader spectrum including abnormal organs, potential lesions, and more generally, any anomalies meaningful for clinical decision-making.
A concurrent work~\cite{baharoon2025rexgroundingct} provide segmentation for 14 types of abnormalities, while limited to chest CT.
% \xiaoman{add \cite{baharoon2025rexgroundingct,hamamci2024developing}?}

\vspace{.1cm}
\noindent \textbf{Medical image report generation} aims to faithfully interpret the medical images with formal text, which is indispensable for numerous clinic procedures~\cite{rao2025multimodal}. 
Existing datasets with image-report pairs are limited to specific modalities and regions, such as chest X-ray~\cite{johnson2019mimic}, brain MRI~\cite{lei2024autorg} and chest CT~\cite{hamamci2024developing,chen2024bimcv}, overlooking the demands for multi-plane whole-body CT report generation.
Recent progress~\cite{zhang2024radgenome,chen2025large,lei2024autorg,chen2024vision} on grounded report generation is limited to organ-level segmentation and description, failing to localize and interpret fine-grained abnormality.

\vspace{.1cm}
\noindent \textbf{Large visual-language models (LVLM) for medicine} are built on large-scale multi-modal medical data, demonstrating exceptional performance and generalization capabilities across diverse tasks~\cite{wu2023towards,moor2023med,zhang2024generalist,li2023llava,meddr}. Despite their potential application to multi-plane whole-body CT images, most existing approaches fail to provide grounded evidence during generation simultaneously, leading to poor explainability and hallucinations that are more difficult to detect.
\section{Taxonomy for Abnormal Findings}
\label{sec:taxonomy}

Given the diversity, complex relationship, and synonyms of abnormalities,
we first construct a taxonomy system to categorize all the abnormalities into representative classes. 

Our taxonomy is guided by four fundamental principles to ensure scientific rigor, 
systematic organization, and clinical utility:
(i) \textbf{comprehensive coverage.} It should include all the clinically significant abnormalities, encompassing both lesions and the abnormal changes on anatomical structures;
(ii) \textbf{image-based definition.} Each abnormal finding is precisely characterized by its observable features on the radiology image, such as physical properties and morphological characteristics that form the essential foundation for diagnosis;
(iii) \textbf{hierarchical organization.} Findings are grouped by anatomical structures and major systems or organs, aligning with standard image interpretation protocols. While cross-regional structures (like skeletal systems) remain independent groups for conciseness;
(iv) \textbf{modular design.} Each category represents a fundamental, independent finding that can be combined with others to express complex and multifaceted abnormalities.

The taxonomy is developed by 7 senior radiologists with at least 10-year experience from 3 centers, following three stages:
(i) an anatomical hierarchy framework is establish based on standard atlases and authoritative human anatomy textbooks~\cite{gray2008gray}; 
(ii) abnormal findings are systematically cataloged under each anatomical structure, based on the clinical experience and authoritative textbooks~\cite{adam2020grainger,haaga2016ct,Mandell2013CoreRadiology} on radiological imaging and pathology; 
(iii) the radiologists conduct rigorous cross-validation and discussion to establish precise consensus on the definitions of controversial or ambiguous abnormalities. 
We finally identified 404 representative abnormalities across 82 anatomies throughout the human body, organized within 40 major organs or systems.
% The detailed taxonomy is in the Appendix.
% To facilitate evaluation of abnormality descriptions, the radiologists further identify key radiological characteristics (e.g., location, morphology, size, density) that should be documented for each finding. Section~\ref{sec:experiment_settings} will introduce how these criteria are used for report evaluation.
\section{\dataset~}
\label{sec:data}

\begin{figure*}[t]
  \centering
  \includegraphics[width=\textwidth]{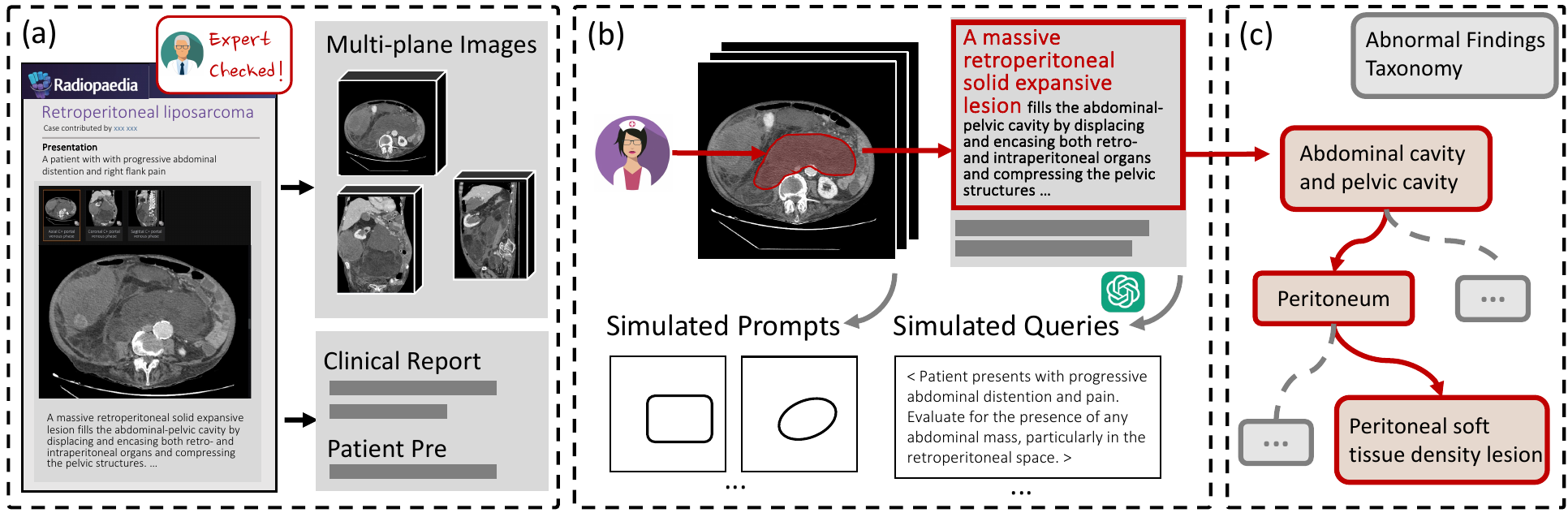} \\
  \vspace{-3pt}
  \caption{\textbf{Data curation overview}. (a) The image-report pairs are collected from an open-sourced and expert-checked website; (b) and (c) Radiologists provide grounding annotation on any abnormalities, link to their text description in reports, and categorize into the taxonomy devised by senior radiologists. The annotation is further extended to instruction data with simulated visual prompts and text queries.
  }
  \vspace{-.3cm}
 \label{fig:data}
\end{figure*}

\subsection{Data Source}
\label{sec:data_source}

Following prior works~\cite{wu2023towards,zheng2024large,wu2024mrgen}, 
we collect raw data from Radiopaedia~\cite{radiopaedia}, a publicly accessible platform of peer-reviewed clinical cases contributed by clinicians worldwide, with appropriate privacy safeguards. 
After application, we have obtained proper permission for all data used in this study. 
Shown in Figure~\ref{fig:data}(a), each case includes radiology images across various planes and anatomical regions, along with patient background information and clinical reports describing findings and diagnoses. 
In total, we collected 46,721 CT images paired with 14,920 clinical reports. 

\subsection{Annotation Pipeline}
\label{sec:annotation_pipeline}

As illustrated in Figure~\ref{fig:data}(b) and (c), 
our goal is to produce grounding annotations for all abnormal findings in the collected CT images,
{\em i.e.}, linking them to corresponding textual descriptions in clinical reports, 
and categorizing them using our proposed taxonomy. 
To reduce annotation burden, we adopt a two pre-processing strategy:
(i) \textbf{image-side pre-processing}. Rather than annotating entire scans, we focus on one key slice and its 8 adjacent slices, selected by original uploading radiologists to be the most informative and representative; 
(ii) \textbf{text-side pre-processing}. We employ GPT-4o~\cite{gpt4o} to extract findings from the reports and identify abnormalities using RaTEScore~\cite{zhao2024ratescore}.

Then, we recruited 4 board-certified radiologists as annotators through a selection phase: candidates each annotated 100 test cases; senior radiologists then selected the top four for full-scale annotation.
To ensure high-quality annotations, we followed three core principles:
(i) \textbf{representative slices}. Select the most representative slice(s) for annotation when abnormalities are present;
(ii) \textbf{complete annotation}. Annotate all abnormal findings visible on selected slices;
(iii) \textbf{vision-language consistency}. Only annotate abnormalities identifiable in provided slices; similarly, linked descriptions must reflect findings visible in these slices. 
For example, when a report mentions organ enlargement, the finding should not be annotated if enlargement cannot be confidently determined from the available slices.

Throughout the annotation, senior radiologists provided real-time consultation and guidance to resolve uncertain cases from the annotators.
Meanwhile, we continuously refined \dataset~ based on annotator feedback: 
(i) \textbf{taxonomy refinement}. We remove or merge 4 less representative categories, 
add 22 previously omitted but clinically important ones, and revise 21 categories for improved clarity and expression;
(ii) \textbf{long-tail mitigation}. To address data imbalance, we identify the underrepresented categories and use GPT-4o~\cite{gpt4o} to filter scans referencing these rare findings, prioritizing them for annotation. 
More details about the long-tail mitigation can be found in the Appendix.

\subsection{Instruction Data Construction}
\label{sec:instruction_generation}

As shown in Figure~\ref{fig:data}(b), we extend our annotated data to support three instruction tasks that simulate realistic clinical scenarios:
(i) \textbf{grounded report generation} simulates a comprehensive examination. We format instructions to detect, ground, and describe all abnormal findings on CT images;
(ii) \textbf{text-guided grounded report generation} involves responding to text queries about specific abnormalities of interest. For instance, for a patient with tuberculous empyema history, checking any abnormality related to it. We generate these queries using GPT-4o based on clinical reports, patient presentations, and medical history, simulating how radiologists approach images with prior knowledge;
(iii) \textbf{visual prompted report generation} focuses on interpreting the marked abnormality. We create visual prompts from annotation masks using various formats, 
{\em e.g.}, bounding box, ellipse, contour, and cropped region.
This simulates a semi-automated workflow where the model elaborates on the clinician-identified abnormality in detail. 
More details about the instruction data are in the Appendix.

\subsection{Summary}
\label{sec:summary}

We prioritize axial images for annotation, while also incorporate coronal and sagittal images based on clinical recommendations. 
In total, we annotated 18,969 abnormalities on 9,990 axial images, 2,738 coronal images, and 1,803 sagittal images. 
Representative examples and the distribution of annotated abnormalities are shown in Figure~\ref{fig:teaser}.
Annotation quality and inter-annotator consistency is verified by a radiologist with 12-year experience, detailed in the Appendix.
\section{\model~}
\label{sec:model}

% This section presents the details of \model~, 
% starting with the problem formulation in Section~\ref{sec:problem_formulation}, then its architectural details in Section~\ref{sec:archtecture} and finally the training details in Section~\ref{sec:training_details}.

This section presents the details of \model~, a novel system for grounded CT image interpretation .
We start with the problem formulation in Section~\ref{sec:problem_formulation}, then its architectural details in Section~\ref{sec:archtecture} and finally the training details in Section~\ref{sec:training_details}.

\begin{figure*}[t]
  \centering
  \includegraphics[width=\textwidth]{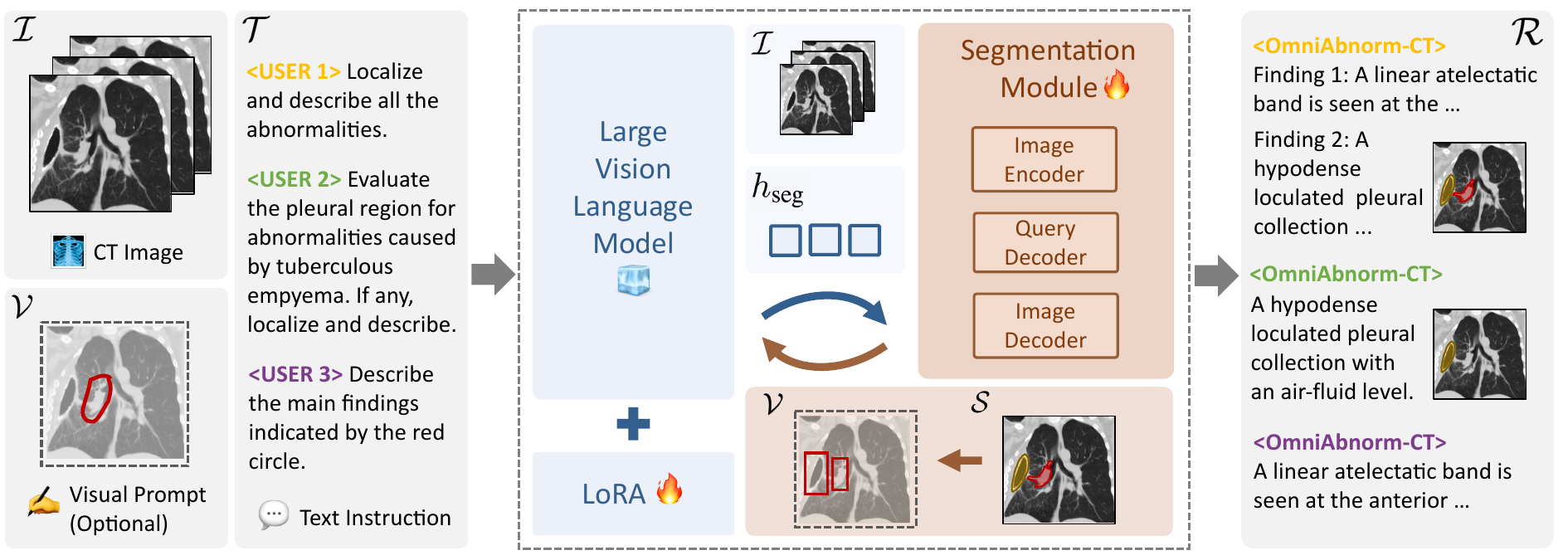} \\
  \vspace{-3pt}
  \caption{\textbf{\model~}. We bridge a VLM and a segmentation module, to allow grounding evidence acquisition during the generation of abnormality description, and further enhance its comprehension for flexible usage with text instruction and visual prompts.}
 \label{fig:model}
 \vspace{-.3cm}
\end{figure*}

\subsection{Problem Formulation}
\label{sec:problem_formulation}
Given a CT image~($\mathcal{I} \in \mathbb{R}^{H \times W \times D}$), 
we aims to build a model $\Phi_\theta(\cdot)$,
that generates the textual descriptions for abnormalities, 
with intermediate groundings when necessary:
\begin{equation}
\{\mathcal{R}, \mathcal{S}\} = \Phi_\theta(\mathcal{T}, \mathcal{I}, \mathcal{V}),
\end{equation}
where $\mathcal{T}$ is the text instruction, $\mathcal{V} \in \mathbb{R}^{H \times W}$ (optional) is the visual prompt on a key slice from the users, for example, bounding boxes, ellipse, contour. $\mathcal{R}$ denotes the generated description for abnormalities, and $\mathcal{S} \in \mathbb{R}^{H \times W}$ refers to the (optional) grounding results on the key slice.

In contrast to the conventional approaches, $\Phi_\theta(\cdot)$ can interpret user-provided textual queries and ground arbitrary abnormal findings across multi-plane CT images spanning the entire human body, and generate corresponding descriptions. 
Additionally, it also supports visual prompts that allow the model to elaborate on user-highlighted abnormalities, enabling an interactive workflow where users can iteratively refine or correct the grounding results for more precise and context-aware descriptions.

\begin{table*}[!t]
\centering
\small
% \resizebox{0.9\textwidth}{!}{%
\setlength{\tabcolsep}{4pt}
\begin{tabular}{l|cccccccccc}
\toprule
\textbf{Model} & \textbf{Rubric} & \textbf{BLEU-1} & \textbf{BLEU-2} & \textbf{BLEU-3} & \textbf{RaTESc} & \textbf{BERTSc} & \textbf{METEOR} & \textbf{ROUGE-1} & \textbf{ROUGE-L} & \textbf{RadG} \\
\midrule
\rowcolor{gray!20} \multicolumn{11}{c}{\textit{\textbf{Axial (n=2193)}}} \\
\midrule

GPT4o & \underline{33.41} & \underline{15.90} & \underline{6.22} & \underline{1.63} & 39.04 & 84.57 & \textbf{19.61} & \underline{18.78} & 14.55 & \underline{6.52} \\
QWen25VL7B(OG) & 7.14 & 5.19 & 0.98 & 0.13 & 14.38 & 37.14 & 5.86 & 5.92 & 4.47 & 0.72 \\
ViP-LLaVA & 8.31 & 15.24 & 4.50 & 0.63 & 33.07 & \underline{85.12} & 13.67 & 17.49 & \underline{14.69} & 3.28 \\
MedDr \checkmark & 20.82 & 9.43 & 2.35 & 0.33 & 32.27 & 78.62 & 9.01 & 12.19 & 10.29 & 3.67 \\
LLaVA-Med \checkmark & 15.41 & 15.62 & 4.25 & 0.55 & \underline{39.11} & 84.65 & 15.72 & 17.76 & 13.60 & 6.13 \\
BiomedGPT \checkmark & 20.37 & 10.85 & 4.25 & 0.72 & 38.29 & 83.18 & 15.88 & 13.50 & 10.91 & 4.64 \\
\rowcolor{lightskyblue!20} \model~ \checkmark & \textbf{35.69} & \textbf{18.03} & \textbf{6.65} & \textbf{1.80} & \textbf{42.81} & \textbf{86.35} & \underline{19.40} & \textbf{21.66} & \textbf{17.61} & \textbf{9.98} \\

\midrule
\rowcolor{gray!20} \multicolumn{11}{c}{\textit{\textbf{Coronal (n=750)}}} \\
\midrule

GPT4o & 25.41 & 9.25 & 3.12 & \underline{0.92} & 27.26 & 59.17 & 9.25 & 12.08 & 9.90 & 4.26 \\
QWen25VL7B(OG) & 26.25 & 13.43 & 3.43 & 0.65 & 37.34 & 84.19 & 13.34 & 14.73 & 11.27 & 3.34 \\
ViP-LLaVA & 10.27 & 13.69 & 3.67 & 0.51 & 35.08 & \underline{84.58} & 12.54 & 16.91 & \underline{13.88} & 3.68 \\
MedDr \checkmark & \underline{26.79} & 8.22 & 2.03 & 0.26 & 32.98 & 76.53 & 8.17 & 11.71 & 9.81 & 4.13 \\
LLaVA-Med \checkmark & 22.34 & \underline{15.60} & 4.11 & 0.65 & \underline{40.96} & 84.39 & 14.15 & \underline{17.99} & 13.45 & \underline{6.46} \\
BiomedGPT \checkmark & 20.25 & 12.58 & \underline{4.79} & 0.74 & 38.69 & 82.90 & \underline{15.37} & 14.69 & 11.46 & 4.39 \\
\rowcolor{lightskyblue!20} \model~ \checkmark & \textbf{31.30} & \textbf{18.38} & \textbf{7.19} & \textbf{2.39} & \textbf{42.88} & \textbf{86.00} & \textbf{19.00} & \textbf{21.30} & \textbf{16.87} & \textbf{9.44} \\

\midrule
\rowcolor{gray!20} \multicolumn{11}{c}{\textit{\textbf{Sagittal (n=591)}}} \\
\midrule

GPT4o & \underline{24.39} & 10.23 & 3.45 & \underline{0.88} & 32.41 & 72.53 & 10.01 & 13.77 & 11.65 & 4.00 \\
QWen25VL7B(OG) & 20.09 & 13.22 & 3.47 & 0.59 & 35.51 & 83.93 & 12.82 & 14.54 & 11.23 & 2.95 \\
ViP-LLaVA & 11.56 & 12.87 & 3.81 & 0.38 & 34.24 & \underline{84.39} & 11.86 & 16.19 & 13.44 & 2.84 \\
MedDr \checkmark & 19.79 & 6.89 & 1.70 & 0.26 & 30.32 & 74.72 & 6.91 & 10.07 & 8.37 & 2.29 \\
LLaVA-Med \checkmark & 15.68 & \underline{15.01} & 4.66 & 0.69 & \underline{38.97} & 84.06 & 14.13 & \underline{17.64} & \underline{13.46} & \underline{5.09} \\
BiomedGPT \checkmark & 17.59 & 12.37 & \underline{4.73} & 0.70 & 36.24 & 82.54 & \underline{14.93} & 14.35 & 11.22 & 3.12 \\
\rowcolor{lightskyblue!20} \model~ \checkmark & \textbf{29.00} & \textbf{17.29} & \textbf{6.48} & \textbf{1.82} & \textbf{41.77} & \textbf{85.62} & \textbf{17.45} & \textbf{20.43} & \textbf{15.87} & \textbf{7.51} \\

\bottomrule
\end{tabular}%
% }
\caption{\textbf{Quantitative results on visual prompted report generation.} Results are averaged within each category and then across all categories, with the best \textbf{bolded} and the second best \underline{underlined}. Models optimized with medical data are marked with \checkmark.
}
\label{tab:visual_prompted_generation}
\vspace{-6pt}
\end{table*}

\subsection{Architecture}
\label{sec:archtecture}

The overall architecture of \model~ is illustrated in Figure~\ref{fig:model}, 
which comprises a VLM and a segmentation module. 
Our core innovation lies in the integration of them, which enables localizing the abnormality and generating a textual description~(findings), supporting flexible interaction through both visual prompts and textual instructions.

\vspace{.1cm}
\noindent \textbf{Vision language model} functions as the core to comprehend user instruction, invoke the segmentation module properly, 
and document the abnormal findings based on the grounding evidence.
Specifically, in one workflow to generate grounded reports, given a CT image~($\mathcal{I}$) and text instruction~($\mathcal{T}$) as user input, 
it will first reason for the specific abnormal findings of interest for the user (or simply any abnormalities). 
For instance, when referring to patient history of tuberculous empyema, it should focus on related abnormalities including pleural thickening, calcifications, etc.
% \weidi{the above sentence is unclear, I don't understand what does it mean by `reason for the specific abnormal findings', can you show me some examples of the text instruction.} \ziheng{Ok.}

Then, to let it acquire grounding evidence, we expand the vocabulary with special token \texttt{<SEG>} as grounding request, which will invoke the segmentation module during generation. We extract its hidden state $h_{\text{seg}}$ from the last decoder layer as the prompt for the segmentation module, 
which encapsulates the targeted abnormal findings:
\begin{equation}
\mathcal{S} = \Phi_{\text{seg}}(\mathcal{I}, h_{\text{seg}}), \text{ } h_{\text{seg}} \in \mathbb{R}^{d}
\label{eq:derive_segmentation}
\end{equation}
Given the segmentation results~($S$), 
we convert it to visual prompt~($\mathcal{V}$) in box format, 
% \weidi{Is this visual prompt a mask ?} \ziheng{Clarified.}
{\em i.e.}, directly overlay it on the original image to create a new input image~($\mathcal{I}'$), and the VLM could seamlessly continue generating the abnormality description~($\mathcal{R}$) based on it:
\begin{equation}
\mathcal{R} = \Phi_{\text{MLLM}}(\mathcal{T},\mathcal{I}'), \text{ } \mathcal{I}' = \mathcal{I} \oplus \mathcal{V}
\label{eq:vis_prmpt_gen}
\end{equation}
where $\oplus$ represents superimposition. This leverages the inherent ability of VLM to perceive visual markers~\cite{cai2024vip}, guiding its attention to specific regions of interest. 

Note that, the above design naturally supports another workflow when users have manually delineated specific abnormalities, or intend to refine the grounding results. 
In such case, $\mathcal{V}$ can also be user-input visual prompts on the image, and the VLM elaborates the user-highlighted abnormality following Equation~\ref{eq:vis_prmpt_gen}. 

\vspace{.1cm}
\noindent \textbf{Segmentation module} is invoked for abnormality grounding based on the CT image~($\mathcal{I}$) and prompt~($h_\text{seg}$), summarized as Equation~\ref{eq:derive_segmentation}. 
Specifically, it first adopts an encoder-decoder backbone to derive image features:
\begin{align}
    (v, u) = \Phi_{\text{seg}}(\mathcal{I}), \text{ } v \in \mathbb{R}^{H' \times W' \times d'}, \text{ } u \in \mathbb{R}^{H \times W \times d},
\end{align}
where $v$ denotes concatenation of multi-scale image embeddings along the channel dimension, which are down-sampled from each encoder layer to a unified resolution,
and $u$ is the pixel-level dense feature from the last decoder layer. To bridge the gap between latent spaces, 
we align the segmentation prompt~($h_\text{seg}$) with the image embeddings~($v$):
\begin{equation}
    q = \Phi_{\text{crossattn}}(v, f(h_\text{seg})), \text{ } q \in \mathbb{R}^{d}
\end{equation}
where $f$ is a projection layer for dimension consistency, and $\Phi_{\text{crossattn}}$ is a cross-attention module treating $v$ as key and value,  $f(h_\text{seg})$ as query. The segmentation prediction is then derived by performing dot product between the adapted feature $q$ and $u$:
\begin{equation}
    \mathcal{S} = q \cdot u, \text{ }  \mathcal{S} \in \mathbb{R}^{H \times W},
\end{equation}

\subsection{Training}
\label{sec:training_details}

\noindent \textbf{Training object.} 
We train the VLM and the segmentation module jointly with a text generation loss and a segmentation loss:
\begin{align}
    \mathcal{L}_{\textbf{txt}} & = \text{CE}(\hat{\mathcal{R}}, \mathcal{R}), \\
    \mathcal{L}_{\textbf{seg}} &= \text{BCE}(\hat{\mathcal{S}}, \mathcal{S}) + \text{DICE}(\hat{\mathcal{S}}, \mathcal{S})
    \label{equ:loss}
\end{align} 

\vspace{.1cm}
\noindent \textbf{Training data.} 
Our training data primarily originates from \dataset~, 
and is formulated into three instruction data formats as illustrated in Section~\ref{sec:instruction_generation}.
Additionally, we incorporate two types of data as supplements: 
(i) \textbf{lesion segmentation data}. This type of data can be formulated into grounding tasks and utilized to enhance the grounding ability. 
A detailed list of these datasets can be found in the Appendix;
(ii) \textbf{medical visual question-answering (VQA) data}. This type of data is involved to maintain our model's generalization capabilities with more diverse question types. 
Specifically, we take CT images and corresponding question-answering data from PubMedVision~\cite{chen2024huatuogpt}.

\vspace{.1cm}
\noindent \textbf{Implementation details.}
In practice, we adopt QWen2.5-VL-7B~\cite{yang2024qwen2} as the VLM, 
and a 6-layer U-Net as the segmentation module. The cross-attention module consists of 6 layers.
We feed 9 consecutive slices to the segmentation module, matching the annotation setup.
Our collaborating radiologists confirm this window provides sufficient local 3D context for the vast majority of clinically significant findings.
However, as detailed in the Appendix, expanding the input context (i.e., multiple slices) does not improve VLM report generation. 
We therefore feed only the center slice to the VLM.
To save computational cost, we only optimize the inserted LoRA layers~\cite{hu2022lora} in the VLM while leaving the whole segmentation module trainable.
More details, runtime and computational analysis are in the Appendix.
\section{Experiments}
\label{sec:experiment}

\begin{table*}[t]
\centering
\small
% \resizebox{\textwidth}{!}{%
\begin{tabular}{l|c|cccccccccc}
\toprule
\textbf{Model} & \textbf{DSC} & \textbf{Rubric} & \textbf{B-1} & \textbf{B-2} & \textbf{B-3} & \textbf{RaTESc} & \textbf{BERTSc} & \textbf{MTR} & \textbf{R-1} & \textbf{R-L} & \textbf{RadG} \\
\midrule
\rowcolor{gray!20} \multicolumn{12}{c}{\textit{\textbf{Axial (n=2193)}}} \\
\midrule

MedULS + LLaVA-Med & 14.92 & 3.00 & 10.60 & 5.31 & 0.50 & 26.84 & 72.92 & 15.90 & 12.46 & 10.50 & 2.26 \\
LiSA + LLaVA-Med & \underline{20.62} & \underline{4.87} & 12.85 & 6.53 & 0.70 & 31.09 & 81.68 & 19.11 & 14.46 & 12.32 & 2.75 \\
BiomedParse + LLaVA-Med & 15.69 & 3.45 & \underline{13.35} & \underline{6.86} & \underline{0.80} & \underline{31.61} & \underline{83.04} & \underline{19.62} & \underline{14.98} & \underline{12.77} & \underline{2.84} \\
\rowcolor{lightskyblue!20} \model~ \checkmark & \textbf{36.04} & \textbf{13.03} & \textbf{18.50} & \textbf{11.75} & \textbf{7.90} & \textbf{33.24} & \textbf{86.10} & \textbf{21.83} & \textbf{21.65} & \textbf{18.75} & \textbf{3.78} \\

\midrule
\rowcolor{gray!20} \multicolumn{12}{c}{\textit{\textbf{Coronal (n=750)}}} \\
\midrule

MedULS + LLaVA-Med & 12.73 & 4.03 & 9.54 & 4.91 & 0.58 & 28.65 & 72.77 & 13.75 & 12.61 & 10.68 & 2.24 \\
LiSA + LLaVA-Med & \underline{17.61} & 5.66 & 10.76 & 5.54 & \underline{0.68} & \underline{33.04} & \underline{81.89} & 16.04 & 14.20 & 11.91 & 2.94 \\
BiomedParse + LLaVA-Med & 14.77 & \underline{6.53} & \underline{11.94} & \underline{6.07} & 0.67 & 32.35 & 81.29 & \underline{16.73} & \underline{15.18} & \underline{12.64} & \underline{3.06} \\
\rowcolor{lightskyblue!20} \model~ \checkmark & \textbf{31.65} & \textbf{13.29} & \textbf{19.13} & \textbf{11.86} & \textbf{7.81} & \textbf{35.47} & \textbf{85.58} & \textbf{20.66} & \textbf{22.14} & \textbf{18.26} & \textbf{4.55} \\

\midrule
\rowcolor{gray!20} \multicolumn{12}{c}{\textit{\textbf{Sagittal (n=591)}}} \\
\midrule

MedULS + LLaVA-Med & 12.87 & 2.16 & 9.71 & 4.97 & \underline{0.70} & 26.06 & 69.51 & 13.18 & 12.26 & 10.50 & 1.64 \\
LiSA + LLaVA-Med & \underline{14.82} & 4.36 & 12.10 & 6.16 & 0.57 & 31.40 & \underline{82.77} & \underline{16.95} & 14.84 & 12.76 & 2.00 \\
BiomedParse + LLaVA-Med & 13.72 & \underline{4.51} & \underline{12.25} & \underline{6.21} & 0.57 & \underline{31.90} & 81.70 & 16.93 & \underline{14.98} & \underline{12.82} & \underline{2.31} \\
\rowcolor{lightskyblue!20} \model~ \checkmark & \textbf{34.38} & \textbf{10.11} & \textbf{17.10} & \textbf{10.63} & \textbf{7.10} & \textbf{33.13} & \textbf{85.25} & \textbf{19.05} & \textbf{20.92} & \textbf{17.67} & \textbf{3.70} \\

\bottomrule
\end{tabular}%
% }
\caption{\textbf{Quantitative results on grounded report generation.} Results are averaged within each category and across all categories, with the best results \textbf{bolded} and the second best \underline{underlined}.}
\label{tab:grounded_report_generation}
\vspace{-.3cm}
\end{table*}

In this section, we first illustrate the experimental settings in Section~\ref{sec:experiment_settings}, including test data, task formulation, and baselines;
Then we introduce the evaluation metrics for different tasks in Section~\ref{sec:evaluation_metrics}, including existing metrics, and propose AbnormRubric for report generation evaluation;
Finally, we provide detailed internal validation results in Section~\ref{sec:experiment_results}, and external validation results in Section~\ref{sec:external_results}.
Due to limited space, qualitative experiment results, and ablation studies on the segmentation module, context range, and visual prompts are presented in the Appendix.

\subsection{Experiment Settings}
\label{sec:experiment_settings}

% Based on \dataset~, we build a rigorous and comprehensive benchmark to evaluate both grounding and description generation abilities for abnormalities.

\noindent \textbf{Internal validation data.} 
We split \dataset~ into the train and test set for training and internal validation, following three principles:
(i) Axial, coronal, and sagittal images are split independently;
(ii) To guarantee a comprehensive benchmark, we allocate at least 5 samples per category to the axial test set (2 for coronal and sagittal), or all samples if fewer exist, with the remaining data following a 3:1 train-test ratio. 
(iii) To avoid data leakage, images from the same patient are restricted to the same set.

\vspace{.1cm}
\noindent \textbf{External validation data.} 
In addition, we apply the annotation pipeline described in Section~\ref{sec:annotation_pipeline} to 65 chest CT scans sampled from CT-RATE~\cite{hamamci2024developing}, a large-scale non-contrast chest CT dataset collected in Turkey clinical centers, which serves as an external validation set to assess generalizability.
Annotators are instructed to label all visible abnormalities in the full volume and, for each abnormality, to select the most prominent slice ({\em i.e.}, the slice with the largest segmentation area) for evaluation.
This yields a total of 271 abnormality annotations.

\vspace{.1cm}
\noindent \textbf{Task formulation.} 
As detailed in Section~\ref{sec:instruction_generation}, we evaluate methods on three tasks of increasing difficulty: \textbf{visual prompted report generation}, 
\textbf{grounded report generation}, and \textbf{text-guided grounded report generation}. 
For text-guided grounded report generation, we generate queries for both present and absent abnormalities at a 3:1 ratio to simulate realistic diagnostic scenarios.

\vspace{.1cm}
\noindent \textbf{Baseline.} 
For visual prompted report generation, we compare with multi-modal LLMs in both the medical domain (MedDr~\cite{meddr}, BiomedGPT~\cite{zhang2024generalist}, and LLaVA-Med~\cite{li2023llava}) and the general domain (GPT-4o~\cite{gpt4o}, Qwen2.5-VL~\cite{yang2024qwen2}, and ViP-LLaVA~\cite{cai2024vip}).
For (text-guided) grounded report generation, as no such solution exists in the literature, 
we prompt LLaVA-Med to describe abnormal findings based on segmentation results from auxiliary grounding models.
Since the images in \dataset~ lack the original DICOM metadata, 
we only consider 2D segmentation models as baselines, including BiomedParse~\cite{zhao2024foundation}, LiSA~\cite{lai2024lisa}, and MedULS~\cite{de2024uls23}.
More details are in the Appendix. 

\begin{table*}[t]
\centering
\small
% \resizebox{\textwidth}{!}{%
\begin{tabular}{l|c|cccccccccc}
\toprule
\textbf{Model} & \textbf{DSC} & \textbf{Rubric} & \textbf{B-1} & \textbf{B-2} & \textbf{B-3} & \textbf{RaTESc} & \textbf{BERTSc} & \textbf{MTR} & \textbf{R-1} & \textbf{R-L} & \textbf{RadG} \\
\midrule
\rowcolor{gray!20} \multicolumn{12}{c}{\textit{\textbf{Axial (n=2193)}}} \\
\midrule

MedULS + LLaVA-Med & 14.24 & \underline{3.58} & \underline{9.27} & \underline{2.32} & \underline{0.27} & \underline{26.58} & \underline{78.54} & \underline{11.06} & \underline{11.03} & \underline{8.86} & \underline{1.78} \\
LiSA + LLaVA-Med & \underline{17.34} & 0.25 & 1.78 & 0.41 & 0.25 & 4.40 & 22.11 & 2.03 & 2.34 & 2.08 & 0.24 \\
BiomedParse + LLaVA-Med & 16.42 & 0.56 & 1.91 & 0.44 & 0.19 & 4.73 & 23.34 & 2.15 & 2.54 & 2.22 & 0.27 \\
\rowcolor{lightskyblue!20} \model~ \checkmark & \textbf{32.40} & \textbf{14.82} & \textbf{11.94} & \textbf{3.33} & \textbf{0.74} & \textbf{34.80} & \textbf{83.85} & \textbf{12.94} & \textbf{14.65} & \textbf{11.94} & \textbf{4.32} \\

\midrule
\rowcolor{gray!20} \multicolumn{12}{c}{\textit{\textbf{Coronal (n=750)}}} \\
\midrule

MedULS + LLaVA-Med & 12.21 & \underline{3.30} & \underline{10.08} & \underline{2.42} & 0.17 & \underline{27.81} & \underline{78.85} & \underline{10.72} & \underline{11.57} & \underline{9.06} & \underline{1.34} \\
LiSA + LLaVA-Med & 12.45 & 0.10 & 2.18 & 0.48 & \underline{0.24} & 5.05 & 28.04 & 2.47 & 2.99 & 2.67 & 0.27 \\
BiomedParse + LLaVA-Med & \underline{14.41} & 0.29 & 2.25 & 0.45 & 0.22 & 5.46 & 27.92 & 2.49 & 3.02 & 2.65 & 0.27 \\
\rowcolor{lightskyblue!20} \model~ \checkmark & \textbf{27.74} & \textbf{20.29} & \textbf{12.51} & \textbf{4.10} & \textbf{1.22} & \textbf{36.56} & \textbf{82.98} & \textbf{13.61} & \textbf{15.83} & \textbf{12.36} & \textbf{5.72} \\

\midrule
\rowcolor{gray!20} \multicolumn{12}{c}{\textit{\textbf{Sagittal (n=591)}}} \\
\midrule

MedULS + LLaVA-Med & 11.61 & \underline{3.35} & \underline{10.07} & \underline{2.60} & \underline{0.32} & \underline{26.87} & \underline{78.33} & \underline{10.61} & \underline{11.83} & \underline{9.16} & \underline{1.79} \\
LiSA + LLaVA-Med & 11.36 & 0.36 & 2.06 & 0.43 & 0.29 & 5.05 & 24.51 & 2.23 & 2.66 & 2.15 & 0.34 \\
BiomedParse + LLaVA-Med & \underline{14.25} & 0.27 & 2.28 & 0.66 & 0.31 & 5.95 & 29.05 & 2.71 & 3.28 & 2.81 & 0.36 \\
\rowcolor{lightskyblue!20} \model~ \checkmark & \textbf{29.89} & \textbf{18.37} & \textbf{12.80} & \textbf{4.51} & \textbf{1.42} & \textbf{36.04} & \textbf{83.97} & \textbf{14.02} & \textbf{16.20} & \textbf{12.79} & \textbf{4.85} \\

\bottomrule
\end{tabular}%
% }
\caption{\textbf{Quantitative results on text-guided grounded report generation.} Results are averaged within each category and then across all categories, with the best \textbf{bolded} and the second best \underline{underlined}.}
\label{tab:text_refer_grounded_report_generation}
\vspace{-.3cm}
\end{table*}

\subsection{Evaluation Metrics}
\label{sec:evaluation_metrics}

\noindent \textbf{Existing Metrics.} 
For grounding results, we calculate Dice Similarity Coefficient~(DSC).
For generated report, we include natural language generation metrics: BLEU~(B)~\cite{bleu}, BERTScore~(BERTSc)~\cite{zhangbertscore}, METEOR~(MTR)~\cite{banerjee2005meteor}, and ROUGE~(R)~\cite{rouge}; as well as medical-specific metrics: RaTEScore~(RaTESc)~\cite{zhao2024ratescore} and RadGraph~(RadG)~\cite{delbrouck2024radgraph}. 

\vspace{.1cm}
\noindent \textbf{AbnormRubric} 
(Rubric) is proposed as a metric for generated report.
Unlike above metrics that are limited to simple word matching or unexplainable semantic similarity, AbnormRubric aligns with clinical judgment by comprehensively evaluating the abnormalities in the generated report from: 
(i) the detection ratio,
(ii) the description accuracy of radiological attributions, 
and (iii) the hallucination.
Formally, given all the ground-truth abnormalities ($\text{TP}+\text{FN}$), a LLM is first instructed to numerate the detected ($\text{TP}$) and hallucinated abnormalities ($\text{FP}$) in the generated report:
\begin{align}
    & \text{Recall} = \text{TP} / (\text{TP}+\text{FN}) \\
    & \text{Precision} = \text{TP} / (\text{TP} + \text{FP})
\end{align}
% Formally, given all the ground-truth abnormalities ($\text{\#gt}$), a LLM is first instructed to numerate the detected ($\text{\#match}$) and hallucinated abnormalities ($\text{\#hall}$) in the generated report:
% \begin{align}
%     \text{Recall} &= \text{\#match} / \text{\#gt} \\
%     \text{Precision} &= \text{\#match} / (\text{\#match} + \text{\#hall})
% \end{align}
Then, for each detected abnormality, the LLM is instructed to judge whether the generated description align with ground-truth report on different radiological attributions:
\begin{align}
    \text{Accuracy} = \frac{1}{N} \sum_{i=1}^N \mathbb{I}[\text{Attribution}_i \text{ aligns with GT}]
\end{align}
where $\mathbb{I}$ is the indicator function.
\textbf{Notably}, for each type of abnormality in our defined taxonomy, the radiologists have tailored and identified a set of radiological attributions, including location, size, morphology, and density, that are clinically important and should be accurately described.
And the final score is formulated as:
\begin{align}
    &\text{Recall}^{*} = (1 + \text{Accuracy}) / 2 \times \text{Recall} \\
    &\text{Rubric} = 2 \times \frac{\text{Recall}^{*} \times \text{Precision}}{\text{Recall}^{*} + \text{Precision}}
\end{align}
More details of AbnormRubric are in the Appendix.

\subsection{Internal Validation Results}
\label{sec:experiment_results}

\noindent \textbf{Visual Prompted Report Generation.} Table~\ref{tab:visual_prompted_generation} reports that \model~ outperforms all baselines on 29 out of 30 metrics. 
Specifically, compared to the strongest baseline, \model~ achieves average improvements of 2.4 in BLEU-1, 2.8 in RaTEScore and 4.26 in AbnormRubric across all three planes. 
It demonstrates \model~'s effectiveness in semi-automated workflows, 
where radiologists mark the abnormalities and the model provides accurate and detailed interpretations.

\vspace{.1cm}
\noindent \textbf{Grounded Report Generation.} 
As shown in Table~\ref{tab:grounded_report_generation}, \model~ demonstrates superior grounding capabilities, outperforming the best baselines by 15.42, 14.04, and 19.56 in DSC on axial, coronal, and sagittal planes respectively. 
For report generation, \model~ achieves the best performance across all metrics, with notable improvements in AbnormRubric, RadGraph and RaTEScore across all anatomical planes, 
indicating that the generated reports can better align with clinical standards. 
These results demonstrate \model~'s potential for fully automated abnormality localization and report generation in clinical settings.

\vspace{.1cm}
\noindent \textbf{Text-guided Grounded Report Generation.} 
In Table~\ref{tab:text_refer_grounded_report_generation}, \model{} achieves substantially better grounding results, improving DSC over the strongest baseline on axial (+14.24), coronal (+14.14), and sagittal (+14.54).
For report generation, \model{} outperforms in both general overlap metrics (average +1.3 BLEU-1, +2.2 ROUGE-1 across planes) and clinical metrics (average +14.42 AbnormRubric across planes), indicating closer alignment with clinical assessments.
It indicate that \model~ can effectively follows radiologist instructions to ground and interpret the relevant abnormalities.

\begin{table*}[htbp]
\centering
\small
\begin{tabular}{l|c|cccccccccc}
\toprule
\textbf{Model} & \textbf{DSC} & \textbf{Rubric} & \textbf{B-1} & \textbf{B-2} & \textbf{B-3} & \textbf{RaTESc} & \textbf{BERTSc} & \textbf{MTR} & \textbf{R-1} & \textbf{R-L} & \textbf{RadG} \\
\midrule
\rowcolor{gray!20} \multicolumn{12}{c}{\textit{\textbf{Visual prompted report generation (n=271)}}} \\
\midrule
GPT4o & - & \underline{19.15} & 17.37 & \underline{7.59} & \textbf{1.81} & \underline{37.86} & 85.52 & \textbf{20.46} & 19.16 & 15.18 & \underline{3.73} \\
QWen25VL7B(OG) & - & 15.15 & 6.41 & 2.57 & 0.42 & 37.04 & 82.84 & 14.62 & 9.52 & 7.37 & 2.51 \\
ViP-LLaVA & - & 10.31 & 17.59 & 6.20 & 0.55 & 32.88 & 85.46 & 15.33 & \underline{19.18} & \underline{16.36} & 1.95 \\
MedDr \checkmark & - & 13.52 & 3.24 & 1.13 & 0.17 & 20.36 & 54.28 & 3.58 & 4.94 & 4.29 & 0.74 \\
LLaVA-Med \checkmark & - & 12.22 & \underline{17.63} & 7.40 & 1.20 & 34.96 & \underline{85.52} & \underline{19.81} & 19.01 & 15.64 & 2.92 \\
BiomedGPT \checkmark & - & 12.07 & 10.99 & 4.57 & 0.60 & 35.14 & 82.62 & 17.30 & 13.92 & 11.37 & 2.67 \\
\rowcolor{lightskyblue!20} \model~ \checkmark & - & \textbf{24.09} & \textbf{19.09} & \textbf{7.71} & \underline{1.67} & \textbf{39.87} & \textbf{86.64} & 19.46 & \textbf{21.02} & \textbf{17.34} & \textbf{4.27} \\
\midrule
\rowcolor{gray!20} \multicolumn{12}{c}{\textit{\textbf{Grounded report generation (n=271)}}} \\
\midrule
MedULS + LLaVA-Med & 11.63 & 3.96 & \underline{12.86} & \underline{7.09} & \underline{1.32} & 33.53 & 81.88 & \underline{17.31} & 15.88 & 12.87 & 2.50 \\
LiSA + LLaVA-Med & \underline{18.34} & 4.37 & 12.33 & 6.54 & 0.62 & \textbf{35.43} & \underline{83.57} & 17.03 & 15.87 & 12.71 & \underline{2.69} \\
BiomedParse + LLaVA-Med & 11.78 & \underline{5.32} & 11.98 & 6.51 & 0.74 & 34.76 & 83.47 & 16.59 & \underline{16.10} & \underline{13.08} & \textbf{2.87} \\
\rowcolor{lightskyblue!20} \model~ \checkmark & \textbf{22.20} & \textbf{8.62} & \textbf{18.55} & \textbf{12.10} & \textbf{8.14} & \underline{34.93} & \textbf{86.09} & \textbf{21.61} & \textbf{22.36} & \textbf{18.32} & 2.57 \\
\midrule
\rowcolor{gray!20} \multicolumn{12}{c}{\textit{\textbf{Text-guided grounded report generation (n=271)}}} \\
\midrule
MedULS + LLaVA-Med & \underline{11.69} & \underline{3.23} & \underline{9.47} & \underline{2.86} & 0.33 & \underline{25.57} & \underline{74.55} & \underline{11.57} & \underline{10.81} & \underline{8.91} & 1.32 \\
LiSA + LLaVA-Med & 9.72 & 0.17 & 2.85 & 1.43 & 0.89 & 4.68 & 20.26 & 3.08 & 3.32 & 2.90 & 1.40 \\
BiomedParse + LLaVA-Med & 10.50 & 0.33 & 4.33 & 2.06 & \textbf{1.44} & 6.65 & 25.52 & 4.50 & 4.77 & 4.48 & \underline{1.63} \\
\rowcolor{lightskyblue!20} \model~ \checkmark & \textbf{14.13} & \textbf{15.12} & \textbf{12.38} & \textbf{3.76} & \underline{1.18} & \textbf{32.25} & \textbf{76.34} & \textbf{13.55} & \textbf{14.85} & \textbf{11.97} & \textbf{3.55} \\
\bottomrule
\end{tabular}
\caption{\textbf{External validation results across three tasks.}
Metrics are averaged within each abnormality category and then across all categories. 
Within each block, the best result is \textbf{bolded} and the second best is \underline{underlined}. 
For metrics that are not applicable to a given task, we report ``-''. 
Models optimized with medical data are marked with \checkmark.}
\label{tab:ctrate_all_tasks}
\end{table*}

\subsection{External Validation Results}
\label{sec:external_results}

\edited{
As demonstrated in Table~\ref{tab:ctrate_all_tasks},
on the external validation set, \model~ maintains strong performance across all three tasks.
In visual prompted report generation, \model~ achieves the highest AbnormRubric score (+4.94 over the best baseline), indicating that its generated reports align more closely with clinical criteria.
Moreover, in both grounded report generation and text-guided grounded report generation, \model~ surpasses all baselines on report generation metrics and shows superior grounding ability (DSC improvements of +10.57 for grounded report generation and +2.44 for text-guided grounded report generation).
These results suggest that \model~ generalizes well to external datasets.
}

\section{Conclusion}

This paper advances the automatic grounded interpretation of CT imaging from an abnormality-centric view. 
To support this, we develop a hierarchical taxonomy of 404 representative abnormal findings.
We contribute \textbf{\dataset~}, a meticulously annotated dataset with detailed grounding and description annotation for around 19K abnormalities, from 14.5K multi-plane CT images across the entire human body.
Built on this dataset, \textbf{\model} enables automatic grounding and description of abnormalities, while supporting flexible clinical usage with text queries or visual prompts.
We establish three representative tasks and devise a clinically grounded metric to demonstrate \model's superior performance.

\vspace{.3cm}
\noindent \textbf{Limitations.} 
As an early attempt, this paper certainly has limitations and future works to continue: 
On dataset, we annotate representative slices rather than full volumes, primarily due to the high cost of expert labeling. 
We will explore semi-automated tools to scale to full-volume annotations in future.
For \model~, current VLM has limited volumetric perception, constrained by training data and compute resources. 
We plan to curate larger-scale whole-body CT datasets for pretraining and adapting a 3D vision encoder in future. 
Nevertheless, our work pioneers abnormality-centric CT interpretation, improving diagnostic transparency via accurate grounding and detailed characterization of abnormalities. 
We believe this paradigm shift holds substantial potential to transform radiology practice.

% \section{Acknowledgments}

% Weidi would like to acknowledge the funding from Scientific Research Innovation Capability Support Project for Young Faculty~(ZY-GXQNJSKYCXNLZCXM-I22).
% \clearpage
{
    \small
    \bibliographystyle{ieeenat_fullname}
    \bibliography{main}
}
\clearpage
\onecolumn
{
    \centering
    \Large
    \textbf{Rethinking Whole-Body CT Image Interpretation: \\ An Abnormality-Centric Approach}\\
    \vspace{0.5em} Appendix \\
    \vspace{1.0em}
}

\setcounter{page}{1}

\appendix
{
  \hypersetup{linkcolor=black}
  \setlength{\parskip}{2pt}
  \tableofcontents
}

\clearpage

\section{Qualitative Experiment Results}
\label{sec:quality_results}

\begin{figure*}[htbp]
  \centering
  \includegraphics[width=.95\textwidth]{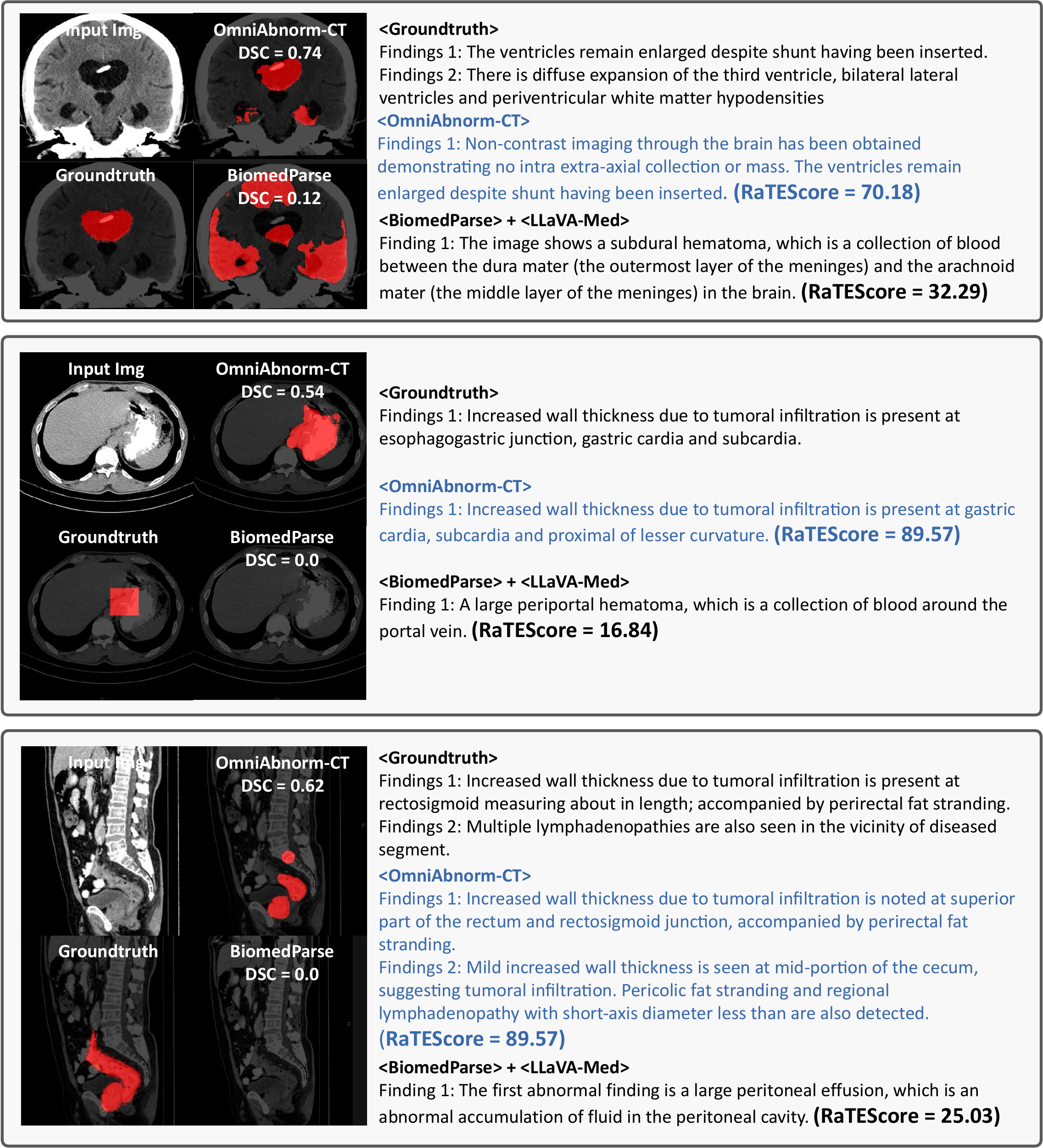} \\
  \caption{Qualitative comparison on the grounded report generation task.
  }
 \label{fig:quality_ground_gen}
 % \vspace{-.3cm}
\end{figure*}

Fig.~\ref{fig:quality_ground_gen} presents several cases from the \textbf{grounded report generation} task, comparing the segmentation and report generation results between \model~ and BiomedParse+LLaVA-Med.
It clearly shows that BiomedParse fails to detect any abnormalities in the latter two cases, and consequently, LLaVA-Med generates reports irrelevant to the abnormalities. 
In contrast, \model~ successfully localizes the abnormalities across all cases and produces more accurate reports.

% \noindent In Fig.~\ref{fig:quality_refer_ground_gen}, we demonstrate several examples from the \textbf{text-guided grounded report generation} task, and compare the segmentation and report generation results of \model~ and BiomedParse+LLaVA-Med.
% As illustrated in the first two examples, BiomedParse fails to accurately localize the queried abnormalities. 
% While in the latter two examples, LLaVA-Med is unable to correctly interpret the segmentation results when generating reports. 
% In contrast, \model~ can successfully localize the queried abnormalities across all examples and generates more precise and clinically relevant reports.

\begin{figure}[htbp]
  \centering
  \includegraphics[width=.95\textwidth]{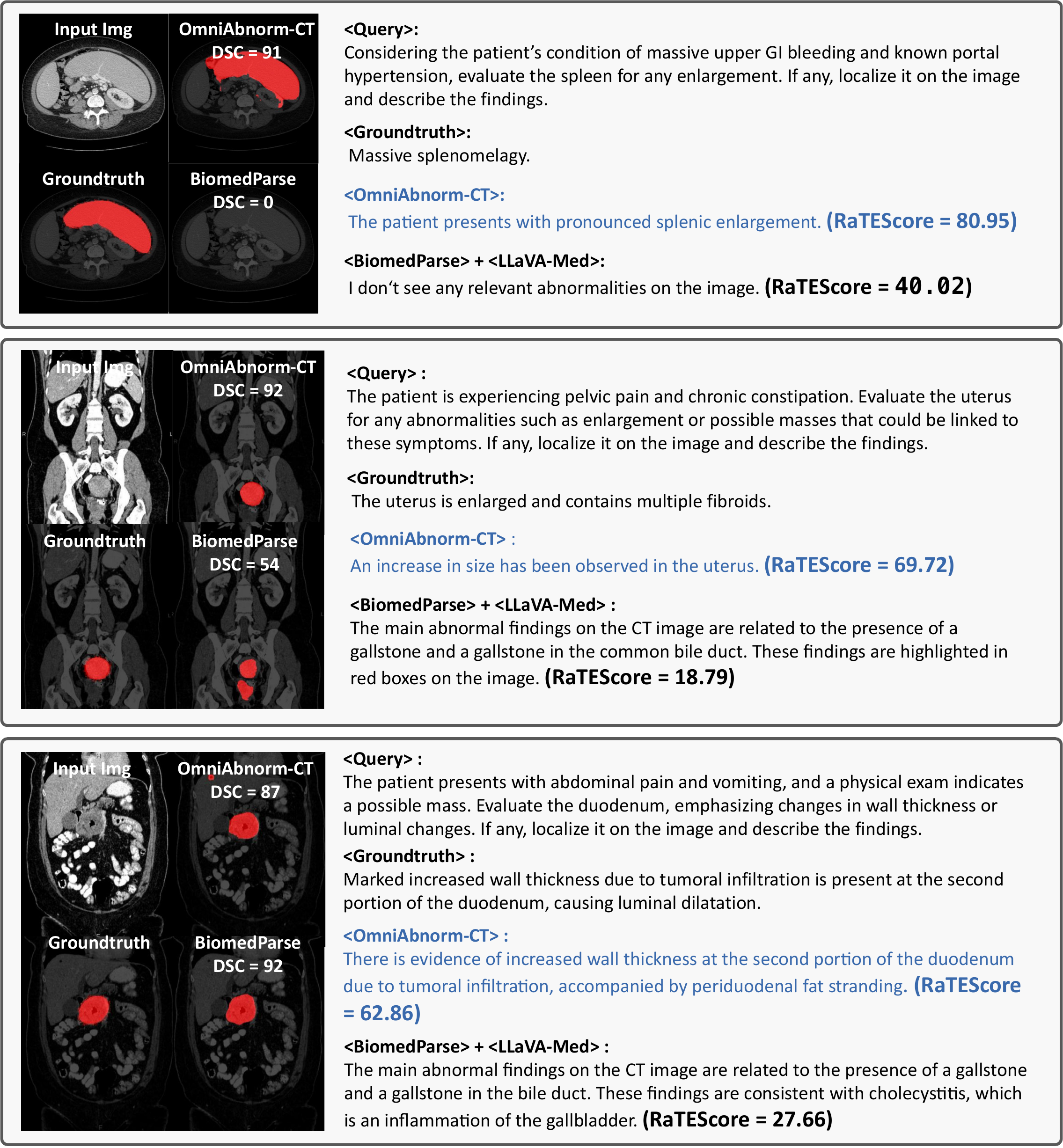} \\
  \caption{Qualitative comparison on the text-guided grounded report generation task.
  }
  % \vspace{.5cm}
 \label{fig:quality_refer_ground_gen}
\end{figure}

\noindent In Fig.~\ref{fig:quality_refer_ground_gen}, we compare \model{} with BiomedParse+LLaVA-Med on the \textbf{text-guided grounded report generation} task. 
BiomedParse mislocalizes the queried findings in first two cases, 
while LLaVA-Med fails to correctly interpret the segmentation results in last two cases. 
In contrast, \model{} consistently localizes the queried abnormalities and produces more precise, clinically aligned reports.

\begin{figure}[htbp]
  \centering
  \includegraphics[width=.95\textwidth]{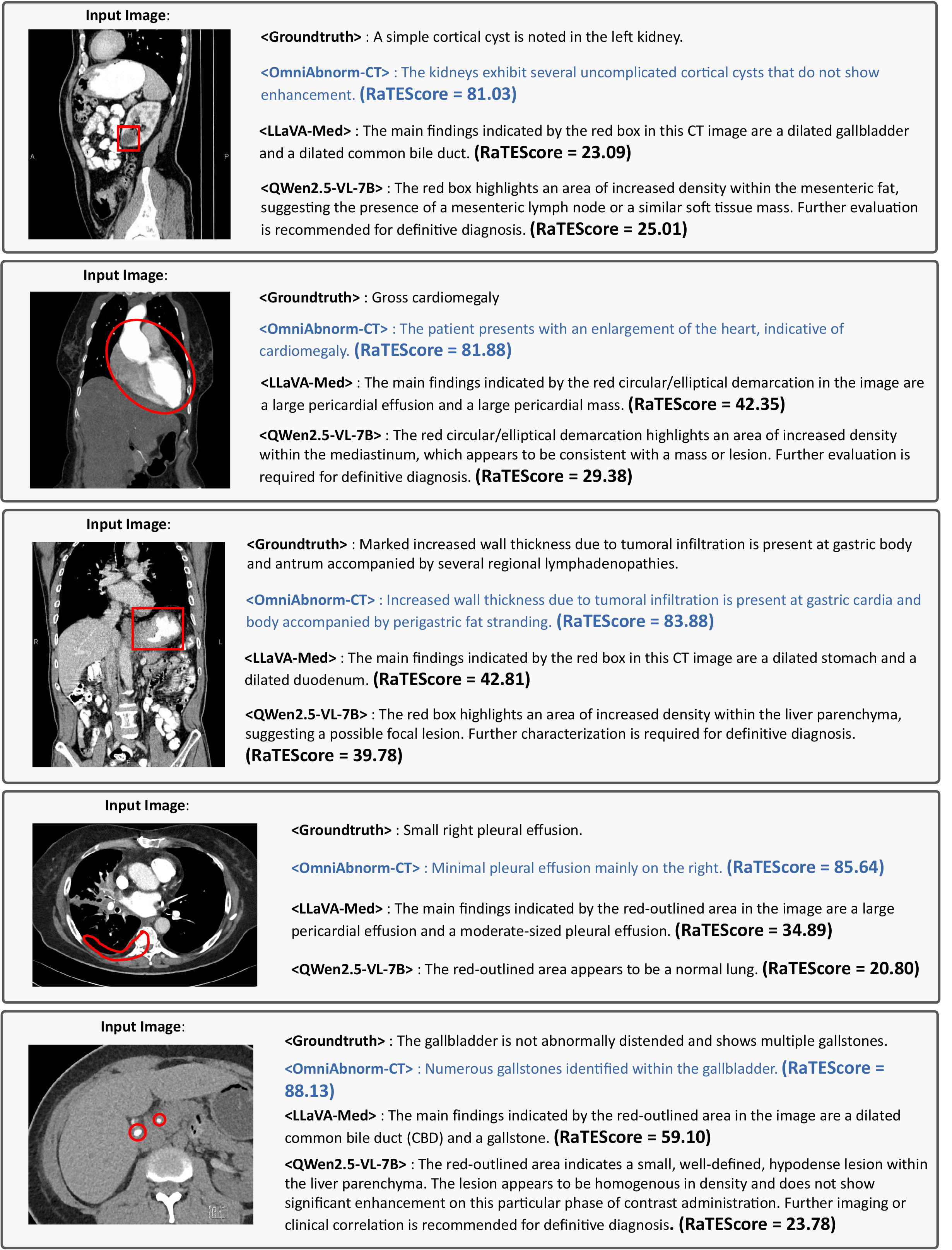} \\
  \caption{Qualitative comparison on the visual prompted report generation task.
  }
 \label{fig:quality_vis_prmpt_gen}
 \vspace{-.3cm}
\end{figure}

\noindent In Fig.~\ref{fig:quality_vis_prmpt_gen}, we compare \model{} with LLaVA-Med and QWen2.5-VL-7B on visual prompted report generation. 
The baselines show some fundamental mistakes, 
{\em e.g.}, LLaVA-Med mislabels the kidney as the gallbladder in the first case,
and QWen2.5-VL-7B confuses the liver and gallbladder in the last case. 
In contrast, \model{} accurately identifies the marked abnormalities and produces higher-quality, clinically consistent reports.

\clearpage

\section{Ablation Studies}
\label{sec:ablation}

We conduct a series of ablation studies on the key factors of \model~, 
from the segmentation module (Section~\ref{sec:ablation_no_seg}), 
the context range of input image (Section~\ref{sec:multislice_context}), 
to the form of visual prompts (Section~\ref{sec:visual_prmpt}).
This section introduces the settings and results of these experiments sequentially. 

\subsection{Ablation Study on Segmentation Module}
\label{sec:ablation_no_seg}

The integration of the segmentation module allows \model~ to acquire grounding evidence for abnormal findings and generate descriptions accordingly. 
To validate its necessity, we conduct an ablation study comparing the full model against a variant without the segmentation module (\model~ w/o Seg), which directly generates findings.

\vspace{.1cm}
\noindent \textbf{Text-guided grounded report generation}. As shown in Table~\ref{tab:ablation_no_seg_refer_ground_gen}, removing the segmentation module significantly degrades performance on most metrics (23/27 across planes).
Furthermore, our manual analysis reveals that, without the grounding evidence, the model struggles to accurately detect the queried abnormalities in CT images.

\vspace{.1cm}
\noindent \textbf{Grounded report generation}. Similarly, Table~\ref{tab:ablation_no_seg_ground_gen} shows that \model{} w/o Seg consistently underperforms the full model on all metrics, reaffirming that segmentation-driven evidence is essential for accurate grounding and generating clinically faithful reports.

\begin{table*}[h]
\centering
\small
% \resizebox{0.8\textwidth}{!}{%
\setlength{\tabcolsep}{5pt}
\begin{tabular}{l|ccccccccc}
\toprule
\textbf{Model} & \textbf{BLEU-1}  & \textbf{BLEU-2}  & \textbf{BLEU-3} & \textbf{RaTESc} & \textbf{BERTSc} & \textbf{METEOR} & \textbf{ROUGE-1} & \textbf{ROUGE-L} & \textbf{RadG} \\
\midrule
\rowcolor{gray!20} \multicolumn{10}{c}{\textit{\textbf{Axial (n=2193)}}} \\
\midrule

\model~ w/o Seg & 4.50 & 2.33 & \textbf{1.21} & 20.53 & 31.06 & 7.04 & 8.50 & 7.02 & \textbf{7.25} \\
\model~ & \textbf{11.94} & \textbf{3.33} & 0.74 & \textbf{34.80} & \textbf{83.85} & \textbf{12.94} & \textbf{14.65} & \textbf{11.94} & 4.32 \\

\midrule
\rowcolor{gray!20} \multicolumn{10}{c}{\textit{\textbf{Coronal (n=750)}}} \\
\midrule

\model~ w/o Seg & 4.50 & 2.47 & 1.27 & 19.07 & 28.34 & 6.27 & 7.96 & 6.43 & \textbf{6.17} \\
\model~ & \textbf{12.51} & \textbf{4.10} & \textbf{1.22} & \textbf{36.56} & 82.98 & \textbf{13.61} & \textbf{15.83} & \textbf{12.36} & 5.72 \\

\midrule
\rowcolor{gray!20} \multicolumn{10}{c}{\textit{\textbf{Sagittal (n=591)}}} \\
\midrule

\model~ w/o Seg & 4.17 & 2.16 & 1.12 & 16.62 & 25.43 & 5.84 & 7.06 & 5.53 & \textbf{5.67} \\
\model~ & \textbf{12.80} & \textbf{4.51} & \textbf{1.42} & \textbf{36.04} & \textbf{83.97} & \textbf{14.02} & \textbf{16.20} & \textbf{12.79} & 4.85 \\

\bottomrule
\end{tabular}%
\vspace{-.1cm}
\caption{Ablation study of the segmentation module for \textbf{text-guided grounded report generation} task. 
% Results are averaged within each category and then across all categories, with the best \textbf{bolded}
}
\label{tab:ablation_no_seg_refer_ground_gen}
\vspace{-.2cm}
\end{table*}

\begin{table*}[h]
\centering
\small
% \resizebox{0.8\textwidth}{!}{%
\setlength{\tabcolsep}{5pt}
\begin{tabular}{l|ccccccccc}
\toprule
\textbf{Model} & \textbf{BLEU-1}  & \textbf{BLEU-2}  & \textbf{BLEU-3} & \textbf{RaTESc} & \textbf{BERTSc} & \textbf{METEOR} & \textbf{ROUGE-1} & \textbf{ROUGE-L} & \textbf{RadG} \\
\midrule
\rowcolor{gray!20} \multicolumn{10}{c}{\textit{\textbf{Axial (n=2193)}}} \\
\midrule

\model~ w/o Seg & 7.73 & 4.74 & 3.16 & 27.87 & 73.14 & 17.65 & 15.18 & 13.58 & 3.42 \\
\model~ & \textbf{19.00} & \textbf{12.26} & \textbf{8.40} & \textbf{33.85} & \textbf{86.24} & \textbf{22.45} & \textbf{22.27} & \textbf{19.43} & \textbf{4.60} \\

\midrule
\rowcolor{gray!20} \multicolumn{10}{c}{\textit{\textbf{Coronal (n=750)}}} \\
\midrule

\model~ w/o Seg & 8.12 & 4.81 & 3.12 & 31.03 & 75.55 & 15.64 & 15.27 & 13.23 & 3.84 \\
\model~ & \textbf{19.28} & \textbf{12.03} & \textbf{7.96} & \textbf{35.40} & \textbf{85.56} & \textbf{20.81} & \textbf{22.35} & \textbf{18.47} & \textbf{4.65} \\

\midrule
\rowcolor{gray!20} \multicolumn{10}{c}{\textit{\textbf{Sagittal (n=591)}}} \\
\midrule

\model~ w/o Seg & 8.79 & 5.21 & 3.33 & 31.21 & 77.41 & 16.18 & 15.86 & 13.87 & \textbf{3.91} \\
\model~ & \textbf{16.69} & \textbf{10.41} & \textbf{6.98} & \textbf{33.45} & \textbf{84.94} & \textbf{18.77} & \textbf{20.67} & \textbf{17.61} & 3.83 \\

\bottomrule
\end{tabular}%
\vspace{-.1cm}
\caption{Ablation study of the segmentation module in \textbf{grounded report generation} task. 
% Results are averaged within each category and then across all categories, with the best \textbf{bolded}
}
\vspace{-.2cm}
\label{tab:ablation_no_seg_ground_gen}
\end{table*}

\subsection{Ablation Study on Context Range}
\label{sec:multislice_context}

In clinical practice, adjacent-slice context is important for CT interpretation, including segmentation and diagnosis~\cite{nnUNET,zheng2024large}. 
To assess its impact on report generation, we vary the number of input slices in visual prompted report generation, using a base QWen2.5-VL-7B. 
Surprisingly, Table~\ref{tab:multi_slice_context} shows that adding slices yields no consistent gains across metrics. 
We hypothesize that current LVLMs, which encode the slices independently into tokens, struggle to model subtle inter-slice spatial relations. 
Accordingly, we use only the center slice as input to the LVLM in subsequent training and evaluation of \model{}.

\begin{table*}[htbp]
\centering
\small
\setlength{\tabcolsep}{5pt}
% \resizebox{0.77\textwidth}{!}{%
\begin{tabular}{l|ccccccccc}
\toprule
\textbf{Input Slices} & \textbf{BLEU-1}  & \textbf{BLEU-2}  & \textbf{BLEU-3} & \textbf{RaTESc} & \textbf{BERTSc} & \textbf{METEOR} & \textbf{ROUGE-1} & \textbf{ROUGE-L} & \textbf{RadGraph} \\
\midrule
\rowcolor{gray!20} \multicolumn{10}{c}{\textit{\textbf{Axial (n=2193)}}} \\
\midrule

4 Adjacent Slices & 9.39 & 2.98 & 0.45 & 31.12 & 83.52 & 14.90 & 11.51 & 9.01 & 2.46 \\
2 Adjacent Slices & 9.24 & 3.03 & 0.46 & 31.47 & 83.49 & 15.05 & 11.41 & 8.99 & 2.49 \\
No Adjacent Slices & \textbf{11.57} & \textbf{4.20} & \textbf{0.76} & \textbf{38.35} & \textbf{84.10} & \textbf{16.68} & \textbf{14.26} & \textbf{11.09} & \textbf{3.85} \\

\midrule
\rowcolor{gray!20} \multicolumn{10}{c}{\textit{\textbf{Coronal (n=750)}}} \\
\midrule

4 Adjacent Slices & 10.93 & \textbf{3.93} & \textbf{0.85} & 32.95 & 83.49 & 15.82 & 13.29 & 9.79 & 3.62 \\
2 Adjacent Slices & 10.89 & 3.85 & 0.73 & 33.06 & 83.48 & \textbf{15.87} & 13.34 & 9.85 & \textbf{3.71} \\
No Adjacent Slices & \textbf{13.62} & 3.54 & 0.66 & \textbf{37.26} & \textbf{84.19} & 13.49 & \textbf{14.97} & \textbf{11.47} & 3.37 \\

\midrule
\rowcolor{gray!20} \multicolumn{10}{c}{\textit{\textbf{Sagittal (n=591)}}} \\
\midrule

4 Adjacent Slices & 9.39 & 2.98 & 0.45 & 31.12 & 83.52 & 14.90 & 11.51 & 9.01 & 2.46 \\
2 Adjacent Slices & 11.48 & \textbf{4.16} & \textbf{0.94} & 33.86 & 83.32 & \textbf{15.70} & 13.64 & 9.97 & \textbf{3.75} \\
No Adjacent Slices & \textbf{13.28} & 3.54 & 0.63 & \textbf{35.67} & \textbf{83.98} & 13.01 & \textbf{14.75} & \textbf{11.34} & 2.91 \\

\bottomrule
\end{tabular}%
\vspace{-.1cm}
\caption{Impact of multi-slice context range on visual prompted report generation, with the base Qwen2.5-VL-7B model as baseline.}
\label{tab:multi_slice_context}
\vspace{-.2cm}
\end{table*}

\subsection{Ablation Study on Different Visual Prompts}
\label{sec:visual_prmpt}

Recent work shows VLMs respond exhibit varying perceptual capabilities for different prompt formats \cite{cai2024vip}. 
We therefore study visual prompt design for visual prompted report generation. 
For each abnormality, we simulate four prompts, {\em i.e.}, center crop, ellipse, contour, and box, and evaluate \model{} separately. 
We also consider a per-case oracle that takes the best score among the four, representing the theoretical best performance via selecting the optimal prompt for each specific abnormality, depending on its shape, location, and etc.
As shown in Table~\ref{tab:ablation_visual_prompt}, cropping performs worst on most metrics (24/27), likely due to lost context. 
Ellipse, contour, and box yield similar performance. 
The oracle notably improves all metrics over any single prompt, revealing large per-case variance. 
This suggests that adaptively choosing the visual prompt by morphology and location is beneficial, whereas cropping away context is harmful.

\begin{table*}[htbp]
\centering
\small
\setlength{\tabcolsep}{4pt}
% \resizebox{0.75\textwidth}{!}{%
\begin{tabular}{l|ccccccccc}
\toprule
\textbf{Prompt Type} & \textbf{BLEU-1}  & \textbf{BLEU-2}  & \textbf{BLEU-3} & \textbf{RaTEScore} & \textbf{BERTScore} & \textbf{METEOR} & \textbf{ROUGE-1} & \textbf{ROUGE-L} & \textbf{RadGraph} \\
\midrule
\rowcolor{gray!20} \multicolumn{10}{c}{\textit{\textbf{Axial (n=2193)}}} \\
\midrule

Center Cropping & 11.70 & 2.87 & 0.68 & 33.03 & 84.84 & 12.77 & 14.58 & 11.75 & 3.92 \\
Ellipse & 12.43 & 3.60 & 1.15 & 35.14 & 84.87 & 13.67 & 15.55 & 12.63 & 5.27 \\
Contour & 12.60 & 3.79 & 1.22 & 35.74 & 85.09 & 14.12 & 15.97 & 12.92 & 5.81 \\
Bounding Box & 12.35 & 3.69 & 1.26 & 35.46 & 84.92 & 13.65 & 15.76 & 12.80 & 5.63 \\
Max & \textbf{18.03} & \textbf{6.65} & \textbf{1.80} & \textbf{42.81} & \textbf{86.35} & \textbf{19.40} & \textbf{21.66} & \textbf{17.61} & \textbf{9.98} \\

\midrule
\rowcolor{gray!20} \multicolumn{10}{c}{\textit{\textbf{Coronal (n=750)}}} \\
\midrule

Center Cropping & 10.94 & 2.96 & 0.80 & 33.57 & 84.48 & 11.92 & 14.23 & 11.36 & 3.87 \\
Ellipse & 12.49 & 3.79 & 1.13 & 36.09 & 84.56 & 13.44 & 15.94 & 12.43 & 5.16 \\
Contour & 12.61 & 3.73 & 1.02 & 35.50 & 84.71 & 13.45 & 15.75 & 12.25 & 4.87 \\
Bounding Box & 12.15 & 3.70 & 1.12 & 35.12 & 84.38 & 13.27 & 15.68 & 12.13 & 4.68 \\
Max & \textbf{18.38} & \textbf{7.19} & \textbf{2.39} & \textbf{42.88} & \textbf{86.00} & \textbf{19.00} & \textbf{21.30} & \textbf{16.87} & \textbf{9.44} \\

\midrule
\rowcolor{gray!20} \multicolumn{10}{c}{\textit{\textbf{Sagittal (n=591)}}} \\
\midrule

Center Cropping & 10.98 & 2.73 & 0.39 & 32.23 & 84.20 & 11.98 & 14.45 & 11.01 & 3.05 \\
Ellipse & 11.37 & 3.10 & 0.65 & 34.36 & 84.01 & 12.19 & 14.45 & 10.95 & 3.46 \\
Contour & 11.08 & 2.95 & 0.57 & 34.19 & 84.22 & 12.01 & 14.38 & 11.10 & 3.22 \\
Bounding Box & 11.29 & 2.98 & 0.67 & 34.07 & 84.13 & 12.16 & 14.84 & 11.24 & 3.45 \\
Max & \textbf{17.29} & \textbf{6.48} & \textbf{1.82} & \textbf{41.77} & \textbf{85.62} & \textbf{17.45} & \textbf{20.43} & \textbf{15.87} & \textbf{7.51} \\

\bottomrule
\end{tabular}%
% }
% \vspace{-.3cm}
\caption{Comparison of different visual prompts in visual prompted report generation task, with \model~ fixed as baseline.
}
\label{tab:ablation_visual_prompt}
\end{table*}

%
%
%

% \section{Limitations \ziheng{TODO: Polish or delete this section}}
% \label{sec:limitations}

% As a pioneering attempt toward automatic grounding and interpretation of abnormalities on multi-plane and whole-body CT images, our work certainly has several limitations: 
% (i) Regarding data annotation, despite providing a large-scale dataset, some rare abnormalities (55 out of 404 in the taxonomy) remain uncovered, as shown in Section~\ref{sec:detailed_distribution}. 
% Additionally, many abnormalities have very few annotations, which significantly restricts the model's capability range. 
% Expanding annotations for these rare categories is treated as our future work. Meanwhile, the CT images in \dataset~ may also exhibit demographic and regional bias.
% (ii) Due to resource constraints, we currently provide grounding annotation for a few representative slices on each CT image, rather than full volumetric segmentation. 
% In the future, we could potentially address this limitation by using interactive segmentation models such as nnInteractive~\cite{isensee2025nninteractive}. 
% (iii) Limited by the 3D image perception capability of multi-modal language models, which is also a common challenge in current generative visual-language foundation models for medicine~\cite{wu2023towards,moor2023med,zhang2024generalist,li2023llava,meddr}, we only considered single slice as input when generating reports. Enhancing models' ability to reason across adjacent slices and integrate volumetric context represents a critical direction for our future research.

%
%
%

\clearpage
\newpage

\section{Details of AbnormRubric}
\label{sec:details_abnormrubric}

\edited{
As introduced in Section~\ref{sec:evaluation_metrics}, the computation of \textbf{AbnormRubric} is decomposed into three stages that explicitly mirror the way radiologists assess a report: (i) abnormality detection (recall), (ii) hallucination checking (precision), and (iii) description accuracy of radiological attributions. All three stages are implemented using LLM-based judgments given the ground-truth report and the generated report.
}

\noindent \textbf{Detection of ground-truth abnormalities (Recall).}
\edited{
Given the set of abnormalities in the ground-truth report, we iterate over each and query an LLM to determine whether this abnormality is explicitly present in the generated report. Abnormalities that are judged as present are counted as true positives (TP), while the remaining ground-truth abnormalities are counted as false negatives (FN). 
The detailed prompt is:
}
{
\begin{tcolorbox}
You are a medical imaging report analyst. 
Your task is to evaluate an AI-generated radiology report (or only a portion/fragment of a report) from two aspects:
\begin{enumerate}
    \item Whether the report is a VALID radiology report (it makes a clear statement about what abnormalities are present OR absent on the current image).
    \item Whether the report identifies a specific abnormality.
\end{enumerate}

\textbf{ABNORMALITY TO CHECK:}
\begin{itemize}
    \item Target Abnormality: \verb|{abnormality_category}|
    \item Relevant Description in Detailed: \verb|{groundtruth_description}|
\end{itemize}

\textbf{AI-GENERATED REPORT TO ANALYZE:}
\verb|{generated_report}|

\textbf{EVALUATION CRITERIA:}
\begin{enumerate}
    \item \textbf{Validity Check:}
    \begin{itemize}
        \item The generated report may contain medically irrelevant content, nonsensical phrases or garbled text. Completely ignore such content and focus only on medically relevant content.
        \item The generated report may be blank, or entirely composed of non-medical gibberish, irrelevant text, or insufficient to determine presence/absence of abnormalities. Then ignore the following procedure and your evaluation is complete: \verb|is_invalid_report=1|, \verb|detected=0|, \verb|description=None|.
    \end{itemize}
    \item \textbf{Detection:}
    \begin{itemize}
        \item Report as \verb|detected=1| if the target abnormality (\verb|{abnormality_category}|) is reported in the generated report.
        \item Consider synonyms and clinically equivalent terms as matches.
        \item Regardless of the description accuracy, such as incorrect location, size measurements, morphological characteristics, density/attenuation, and so on.
        \item The generated report may indicate that no abnormalities were detected on the image, for example, \verb|I don't see any relevant abnormalities on the image.| Then ignore the following procedure and your evaluation is complete: \verb|detected=0|.
    \end{itemize}
    \item \textbf{Description Extraction:}
    \begin{itemize}
        \item If \verb|detected=1|, collect ALL relevant description from the report.
        \item Preserve original wording without modifications.
        \item If multiple descriptions exist, concatenate them.
        \item May include adjacent text if the description cannot be grammatically isolated.
    \end{itemize}
\end{enumerate}

\textbf{REQUIRED OUTPUT FORMAT:}
\begin{verbatim}
{
  "is_invalid_report": 0, // 1 or 0
  "detected": 1,      // 1 if detected, 0 if not
  "description": "exact text" // string if detected, null if not
}
\end{verbatim}
   
\textbf{STRICT PROHIBITIONS:}
\begin{itemize}
    \item NEVER add information not present in the generated report.
    \item NEVER paraphrase or summarize original text.
    \item DO NOT output any additional text except for a valid json.
\end{itemize}
\end{tcolorbox}
}

\noindent \textbf{Identification of hallucinated abnormalities (Precision).}
\edited{
To assess hallucinations, we further instruct the LLM to enumerate any abnormalities that are described in the generated report but do not exist in the ground-truth report.
Abnormalities that cannot be matched are counted as false positives (FP). The detailed prompt is:
}
{
\begin{tcolorbox}

Your task is to evaluate the hallucinations in an AI-generated radiology report (or only a portion/fragment of a report):
\begin{enumerate}
    \item Suppose some ground truth abnormalities have been successfully detected by the AI-generated radiology report.
    \item Identify any \textbf{ADDITIONAL} abnormalities mentioned in the generated report that are \textbf{NOT} covered by the ground truth list.
\end{enumerate}

\textbf{THE AI-GENERATED REPORT TO ANALYZE:} \verb|generated_report|

\textbf{GROUND TRUTH ABNORMALITIES AND MATCHING RESULTS:}
\verb|groundtruth_abnormalities|

\textbf{INSTRUCTIONS:}
\begin{enumerate}
    \item Review matched content: The above shows which ground truth abnormalities were detected and their corresponding text in the generated report.
    \item Find additional abnormalities: Identify ALL ADDITIONAL abnormal findings mentioned in the generated report that are NOT already matched to ground truth abnormalities.
    \item Exclude already matched content: Do NOT include any abnormalities that are already matched to ground truth abnormalities.
    \item Ignore normal findings: Only report abnormalities, not normal findings.
    \item Output: ONLY valid JSON with no additional text.
\end{enumerate}

\textbf{NOTE:}
\begin{enumerate}
    \item The generated report may contain medically irrelevant content, nonsensical phrases or garbled text. Completely ignore such content and focus only on medically relevant content.
    \item If the generated report is entirely composed of non-medical gibberish, irrelevant text, or is blank, then your evaluation is complete: \verb|count = 0| and \verb|hallucinated_abnormalities = []|.
\end{enumerate}

\textbf{REQUIRED OUTPUT FORMAT:}
\begin{verbatim}
{
  "hallucinated_abnormalities": [...],  // list of hallucinated abnormalities
  "count": x  // number of hallucinated abnormalities
}
\end{verbatim}

If no additional abnormalities found:
\begin{verbatim}
{
  "hallucinated_abnormalities": [],
  "count": 0
}
\end{verbatim}

\end{tcolorbox}
}

\noindent \textbf{Accuracy of radiological attributions.}
\edited{
For each \emph{detected} ground-truth abnormality (TP), we assess how accurately its key radiological attributions are described in the generated report. 
Based on our abnormality taxonomy, radiologists define a set of clinically important attributions for each abnormality type (e.g., location, morphology, density, size). 
For a given abnormality, we consider only those attributions that are explicitly mentioned in the ground-truth report and ask the LLM to compare them with the corresponding text in the generated report.  
For each attribution, the LLM returns a binary judgment of whether the generated description is clinically consistent with the ground-truth description.
The detailed prompt is:
}
{
\begin{tcolorbox}
You are a medical imaging report analyst.
For a single abnormality in the image, the AI model has generated a description. 
Your task is to compare the AI-generated description with the ground-truth description and evaluate the AI's accuracy across specific attributes that will be provided.
\textbf{EVALUATION INSTRUCTIONS:}
\textbf{INPUT DATA:}
\begin{itemize}
    \item Abnormality: \verb|{abnormality_category}|
    \item Ground-Truth (GT) Description: \verb|{gt_description}|
    \item AI-Generated (Pred) Description: \verb|{pred_description}|
    \item Attributes: \verb|{attribution_ls}|
\end{itemize}
\end{tcolorbox}
}

{
\begin{tcolorbox}

\begin{enumerate}
    \item \textbf{Detection of Attribute Mention:}
    \begin{itemize}
        \item Evaluate ONLY the pre-defined attributes provided in the list.
        \item For each attribute, check if it is mentioned in the GT description.
        \item For attributes present in the GT description, check if they are also mentioned in the Pred description.
    \end{itemize}

    \item \textbf{Comparison:}
    \begin{itemize}
        \item For attributes present in BOTH GT and Pred:
        \begin{itemize}
            \item Extract the exact phrasing from both descriptions.
            \item Determine if the descriptions are clinically equivalent (1) or not (0).
            \item Consider synonyms and clinically equivalent terms as matches.
            \item Provide brief reasoning for your judgment.
        \end{itemize}
        \item For attributes ONLY present in GT:
        \begin{itemize}
            \item Mark as not clinically equivalent (0).
        \end{itemize}
    \end{itemize}

    \item \textbf{Quantitative Data Handling:}
    \begin{itemize}
        \item If GT contains quantitative data (e.g., measurements with units):
        \begin{itemize}
            \item Extract the numerical value and unit from GT.
            \item Check if Pred provides quantitative data for the same attribute.
            \item If Pred lacks quantitative data: mark as non-equivalent (0).
            \item If Pred provides data: calculate percentage error after unit conversion.
            \item Error $\leq 20\%$: equivalent (1); Error $> 20\%$: non-equivalent (0).
            \item In the reasoning field, include the brief calculation process.
        \end{itemize}
    \end{itemize}

    \item \textbf{Special Cases:}
    \begin{itemize}
        \item If Pred provides quantitative data but GT does not: mark as non-equivalent (0).
        \item For ambiguous cases where clinical equivalence cannot be determined: mark as uncertain ($-1$).
    \end{itemize}

    \item \textbf{Strict Constraints:}
    \begin{itemize}
        \item Base ALL judgments strictly on the provided GT and Pred texts.
        \item NO additions, deletions, modifications, or external inferences.
        \item Preserve exact wording when extracting descriptions.
        \item Clinical relevance overrides literal text matching:
        \begin{itemize}
            \item Ignore differences in writing style, terminology preferences, and phrasing variations.
            \item Consider clinically synonymous terms as equivalent.
            \item Focus on clinical semantic meaning rather than exact word matching.
            \item Account for standard medical abbreviations and their equivalent expressions.
        \end{itemize}
    \end{itemize}
\end{enumerate}

\textbf{REQUIRED OUTPUT FORMAT (JSON only):}
\begin{verbatim}
{
  "attributes": {
    "attribute_name_1": {
      "present_in_gt": 1, // 0 if not present, and the followings can be empty
      "gt_description": "exact text from GT",  // null if not present
      "pred_description": "exact text from Pred",  // null if not present
      "equivalent": 1,  // 0 if not equivalent, -1 if can not be determined
      "reasoning": "brief clinical justification",  // null if not present
      "gt_quantified": 1,  // 0 if not quantified
      "pred_quantified": 1,  // 0 if not quantified
      "error_percentage": 5.2 // null if not both quantified 
      or can not be calculated
    },
    ...
  }
}
\end{verbatim}
\end{tcolorbox}
}

\clearpage

\section{Supplementary Implementation Details}
\label{sec:implement_details}

\subsection{Supplementary Implementation Details of \model~}
\label{sec:implement_details_ours}

\noindent \textbf{Training Hyperparameters.} We adopt 8 NVIDIA A100-80G GPUs for training, with 1 sample per device and gradient accumulation set to 4. We train \model~ for 250K iterations in total, and use 5\% steps for warming up. 
We take AdamW~\cite{loshchilov2017decoupled} as optimizer with a learning rate of 2e-4.
When optimizing the multi-modal language model and the segmentation module jointly, we assign equal weights to the text generation loss and segmentation loss, as defined in Equation~\ref{equ:loss}.
We set the rank of LORA to 32. 

\noindent \textbf{Training Data.} We pad and rescale all images to 512$\times$512. For training data sampling from four tasks: visual prompted report generation, grounded report generation, text-guided grounded report generation, and general VQA, we control the ratio to be 1:1:1:1. Additionally, in the abnormality grounding task, we maintain a ratio of 6:4 between \dataset~ and public lesion segmentation data. 
For each abnormality description, we use the following prompt to have GPT rewrite three different versions, then randomly select one as the ground truth:
{
    \begin{tcolorbox}
    Your task is to produce THREE different rewrites of medical reports while preserving all original medical content and diagnostic information, changing only the expression style and sentence structure. \\
    \\
    Rules: \\
    1. Strictly maintain all medical information, diagnostic results, numerical values, and key findings. \\
    2. Do not add any new medical content or conclusions. \\
    3. Do not remove any medical information from the original report. \\
    4. Only change the wording, sentence structure, word order, and vocabulary choices. \\
    5. Maintain professional medical language and terminology. \\
    6. Preserve the overall structure of the original report (such as section divisions). \\
    7. Ensure all rewritten reports remain medically rigorous and accurate. \\
    8. Create THREE distinct rewrites with different phrasing and structure. \\
    \\
    Input: Original medical report written by a doctor \\
    Output: THREE rewritten versions with identical content but different expression\\
    \\
    Please rewrite the following medical report in three different ways:\\
    \$Findings Description\$\\
    Your response must follow this exact format:\\
    \$Output Template\$
    \end{tcolorbox}
}
\vspace{.2cm}

\subsection{Supplementary Implementation Details of Baselines}
\label{sec:implement_details_baseline}

\textbf{Visual prompted generation.} we prompt each abnormality with all types of visual prompts, and take the maximum score. All the methods are prompted with the following template:
{
    \begin{tcolorbox}
    You are a helpful medical assistant. Describe the abnormal findings indicated by the \$Visual Prompt\$. \\
    Please use precise medical terminology, maintain the concise reporting style used in formal radiology reports and provide only the specific radiological findings. Do not list general possibilities, explanations, or recommendations.
    \end{tcolorbox}
}

\noindent \textbf{Grounded report generation.} 
We integrate LLaVA-Med with 2D segmentation models as baselines. 
Specifically, we re-implement the MedULS~\cite{de2024uls23} with a 2D nnU-Net~\cite{nnUNET} and the public lesion segmentation datasets in Table~\ref{tab:datasets_summary}, covering all the datasets in the official ULS-23 challenge and 10 additional ones.
To our knowledge, this represents the segmentation model with the broadest capability range (in terms of abnormality variety) that can be constructed from currently available public datasets.
For BiomedParse~\cite{zhao2024foundation} and LiSA~\cite{lai2024lisa}, we prompt them with `Abnormal findings on the CT image' to derive segmentation results for all the abnormalities on the input CT image. 
Then, we combine each segmentation model with LLaVA-Med by converting the segmentation results into bounding boxes overlaid on the CT image, and prompt LLaVA-Med to generate the report based on these visual cues:
{
    \begin{tcolorbox}
    You are a helpful medical assistant. The abnormal findings are highlighted in red boxes on this CT image, if present. \\
    Please describe each abnormal finding indicated by the red boxes using the format 'Finding 1: [description]', 'Finding 2: [description]', etc. \\
    Use precise medical terminology, maintain the concise reporting style used in formal radiology reports and provide only the specific radiological findings. Do not list general possibilities, explanations, or recommendations. Respond with 'I don't see any abnormalities on the image.' if no abnormalities are present.
    \end{tcolorbox}
}

\noindent \textbf{Text-guided grounded report generation.} We use the simulated text queries as prompts for BiomedParse and LiSA to derive segmentation results for the queried abnormalities, which are detailed in Section~\ref{sec:instruction_generation}. 
For MedULS, since it doesn't support text-prompted segmentation, we simply use its unconditioned segmentation results.
Similarly, we convert their segmentation results into bounding boxes overlaid on the CT image, and prompt LLaVA-Med to generate the report based on them:
{
    \begin{tcolorbox}
    You are a helpful medical assistant. The abnormal findings are highlighted in red boxes on this CT image, if present. \\
    Please describe the abnormal findings indicated by the red boxes. \\
    Use precise medical terminology, maintain the concise reporting style used in formal radiology reports and provide only the specific radiological findings. Do not list general possibilities, explanations, or recommendations. Respond with 'I don't see any relevant abnormalities on the image.' if no abnormalities are present.
    \end{tcolorbox}
}

\subsection{Runtime and Computation Analysis}

\textbf{Training.} We use 4 A100-80G GPUs for training \model~. 
The segmentation module is first pre-trained for 50,000 steps with batch size 16, which takes around 10 hours. 
Then the Qwen2.5-VL is fine-tuned on the visual-prompted generation task for 20,000 steps, with batch size of 1 and gradient accumulation every 8 steps, which takes around 8 hours.
Finally, the VLM and segmentation modules are jointly trained for grounded report generation, with batch size of 1 and gradient accumulation every 8 steps.
This final stage takes approximately 48 hours for 100,000 steps.

\noindent \textbf{Inference.} All inferences are conducted on 1 A100-80G GPU. 
We use mixed-precision without quantization.
The average inference latency is 0.49 second per sample for visual-prompted generation, 0.98 second per sample for grounded report generation, and 1.01 second per sample for text-guided grounded report generation.

\section{Supplementary Details of \dataset~}

\subsection{Quality Verification}
\label{sec:quality_check}

\noindent \textbf{Annotation Quality.} 
We conducted rigorous and comprehensive quality verification on the annotations in \dataset~. For each annotator, we randomly sampled 100 annotated images on the axial plane, 50 on the coronal, and 50 on the sagittal plane. Then a senior radiologist with 12 years of experience assesses the annotation quality on four key metrics: 
(i) \textbf{Detection rate} measures the percentage of abnormalities successfully identified and annotated by the annotator, without any omission;
(ii) \textbf{Grounding precision} evaluates the percentage of grounding annotations that properly encompass the primary regions of the abnormality, while minimizing false positive areas;
(iii) \textbf{Report concordance} quantifies the percentage of description annotations that faithfully reflect the linked abnormal findings on the image;
(iv) \textbf{Classification accuracy} measures the percentage of abnormalities that are correctly categorized. 
The verification results demonstrated that:
(i) 3 out of 4 annotators achieved perfect scores across all four metrics (100\%). 
(ii) Only one annotator had 10 abnormality annotations where the descriptions were not sufficiently accurate and the category labels were also incorrect, resulting in Report concordance and Classification accuracy of 95\%, while maintaining 100\% Detection rate and Grounding precision. 
These results confirm the high quality of annotations in our \dataset~.

\noindent \textbf{Cross-annotator Consistency.}
We further conducted inter-annotator consistency checks across 4 annotators on 40 random samples. 
For each sample, we enumerated all annotator pairs and, for each pair, computed the DSC between their segmentation masks and the BLEU-1 score between their abnormality-level report descriptions. 
Averaging these metrics across all annotator pairs yielded mean pairwise scores of DSC=84.5 for segmentation and BLEU-1=72.2 for finding descriptions, indicating reasonably good agreement among the annotators.

\subsection{Mitigation of the Long-Tail Distribution}
\label{sec:long_tail_problem}
\begin{figure}[h]
  \centering
  \includegraphics[width=0.9\textwidth]{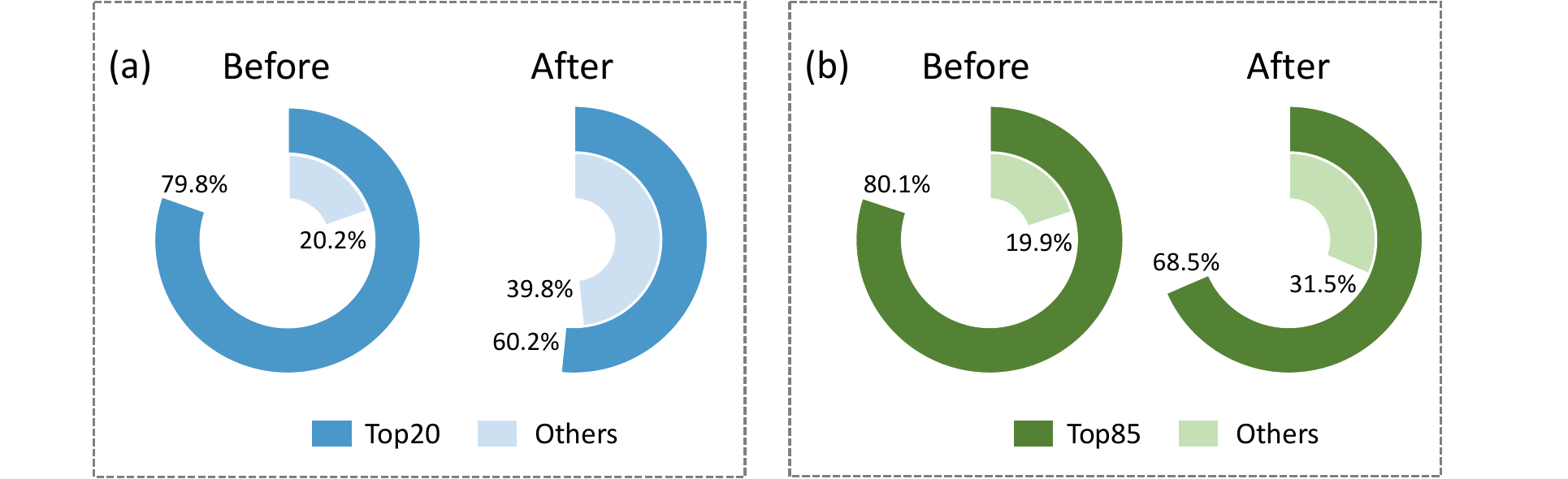} \\
  \caption{The distribution before and after prioritizing annotation for underrepresented organs. (a) Comparison of organ distribution in annotations; (b) Comparison of abnormality category distribution in annotations.
  }
 \label{fig:long_tail}
 \vspace{-.3cm}
\end{figure}

To mitigate the long-tail distribution in \dataset~, 
we identified underrepresented organs in our annotated corpus and strategically employed GPT-4o to analyze unlabeled reports. 
Using the following prompt, we efficiently filtered cases containing abnormal findings related to these underrepresented organs:
{
    \begin{tcolorbox}
    This is a report of a CT scan: \$Report\$ \\
    Please help me carefully check if the report mentions any abnormal findings that belong to the following anatomical areas: \$List of Underrepresented Organs\$. \\
    If there are, please output 'YES', otherwise output 'NO'. Do not output any other information.
    \end{tcolorbox}
}
These identified cases were then prioritized in our annotation pipeline. To evaluate the effectiveness of this strategy, we randomly sampled 1,000 images before and after implementation. 
As shown in Fig.~\ref{fig:long_tail}, prior to our intervention, the top 20 organs (out of 82 total) accounted for 79.8\% of all annotations, while the top 85 abnormality categories (out of 340 total) constituted 80.1\%.
Following our strategy, the underrepresented organs and rare abnormality categories received notably increased annotation coverage: the representation of underrepresented organs increased by 19.6\%, while rare abnormality categories increased by 11.6\%, significantly enhancing the diversity of our dataset.

\clearpage

\subsection{Comparison against Existing Datasets}

We compare \dataset~ with existing public CT image datasets in Table~\ref{tab:datasets_summary}, including those widely used for lesion segmentation, lesion detection, organ segmentation, or report generation. In contrast to the substantial limitations exhibited by these datasets, as detailed the main paper, \dataset represents the first large-scale dataset designed for abnormality grounding and description across multi-plane and whole-body CT images.

\begin{table*}[h]
\centering
\small
\renewcommand{\arraystretch}{1.2}
\setlength{\tabcolsep}{8pt}
\small
% \resizebox{\textwidth}{!}{
\begin{tabular}{lllllll}
\toprule
\textbf{Dataset} & \textbf{Task} & \textbf{Anatomy} & \textbf{Plane} & \textbf{Report} & \textbf{\#Category} & \textbf{\#Image} \\
\midrule
ULS Bone~\cite{de2024uls23} & Lesion Seg. & Bone & Axial & \usym{2717} & 1 & 151 \\
ULS Pancreas~\cite{de2024uls23} & Lesion Seg. & Pancreas & Axial &\usym{2717}& 1 & 119 \\
MSD Liver~\cite{MSD} & Lesion Seg. & Liver & Axial &\usym{2717}& 1 & 131 \\
MSD Lung~\cite{MSD} & Lesion Seg. & Lung & Axial &\usym{2717}& 1 & 63 \\
MSD Colon~\cite{MSD} & Lesion Seg. & Colon & Axial &\usym{2717}& 1 & 126 \\
MSD Pancreas~\cite{MSD} & Lesion Seg. & Pancreas & Axial &\usym{2717}& 1 & 281 \\
COVID19~\cite{COVID19} & Lesion Seg. & Lung & Axial &\usym{2717}& 1 & 20 \\
KiTS23~\cite{KiTS23} & Lesion Seg. & Kidney & Axial &\usym{2717}& 2 & 489 \\
KiPA22~\cite{KiPA22} & Lesion Seg. & Kidney & Axial &\usym{2717}& 2 & 70 \\
NSCLC~\cite{NSCLC} & Lesion Seg. & Lung & Axial &\usym{2717}& 1 & 85 \\
LIDC IDRI~\cite{LIDCIDRI} & Lesion Seg. & Lung & Axial &\usym{2717}& 1 & 750 \\
LNDb~\cite{LNDb} & Lesion Seg. & Lung & Axial &\usym{2717}& 1 & 236 \\
INSTANCE22~\cite{INSTANCE} & Lesion Seg. & Brain & Axial &\usym{2717}& 1 & 100 \\
Seg.Rap2023~\cite{SegRap2023} & Lesion Seg. & Head \& Neck & Axial &\usym{2717}& 2 & 120 \\
FUMPE~\cite{FUMPE} & Lesion Seg. & Lung & Axial &\usym{2717}& 1 & 35 \\
RibFrac~\cite{ribfracchallenge2025} & Lesion Seg. & Rib & Axial &\usym{2717}& 1 & 420 \\
Adrenal ACC Ki67~\cite{AdrenalACCKi67} & Lesion Seg. & Adrenal Gland & Axial &\usym{2717}& 1 & 29 \\
LNQ2023~\cite{dorent2024lnq} & Lesion Seg. & Lung & Axial &\usym{2717}& 1 & 393 \\
NIH-LN~\cite{roth2014new} & Lesion Seg. & Lung & Axial &\usym{2717}& 1 & 175 \\
CCC-18~\cite{CCC18} & Lesion Seg. & Chest \& Abdomen & Axial &\usym{2717}&\usym{2717}& 404 \\
DeepLesion~\cite{yan2018deep} & Lesion Det. & Whole Body & Axial &\usym{2717}& \usym{2717} & 33K \\
TotalSegmentator~\cite{Totalsegmentator} & Organ Seg. & Whole Body & Axial &\usym{2717}&\usym{2717}& 1.2K \\
AbdomenAtlas~\cite{abdomenatlas} & Organ Seg. & Chest \& Abdomen & Axial &\usym{2717}&\usym{2717}& 8K \\
CTRATE~\cite{haaga2016ct} & Report Gen. & Chest & Axial & \checkmark &\usym{2717}& 26K \\
BIMCV-R~\cite{chen2024bimcv} & Report Gen. & Chest & Axial & \checkmark &\usym{2717}& 8K \\
ReXGroundingCT~\cite{baharoon2025rexgroundingct} & Lesion Seg. & Chest & Axial & \checkmark & 14 & 3K \\
\hline
& \textbf{Lesion Seg.} & & \textbf{Axial} & \checkmark & \textbf{340} & \textbf{10K} \\
\textbf{\dataset~} & \textbf{Lesion Det.} & \textbf{Whole Body} & \textbf{Coronal} & \checkmark & \textbf{255} & \textbf{2K} \\
& \textbf{Report Gen.} & & \textbf{Sagittal} & \checkmark & \textbf{223} & \textbf{1.6K} \\
\bottomrule
\end{tabular}
% }
\caption{Comparison of key characteristics between \dataset~ and widely-used public CT imaging datasets. Note that CCC18 and DeepLesion has no category label for each annotated lesion. Even though some scans in these datasets are isotropic, they are acquired as axial-plane. Abbreviations: Category = Abnormality Category; Seg. = Segmentation; Det. = Detection; Gen = Generation.}
\label{tab:datasets_summary}
\end{table*}

\subsection{Generation of Text Queries and Visual Prompts}
\label{sec:txt_refer_vis_prmpt}

We simulate four visual prompts for each annotated abnormality, mimicking how clinicians would select the most appropriate highlighting method based on the abnormality's shape and location:
(i) \textbf{cropped region.} We extract the minimum bounding box that completely contains the annotated lesion region, add a 50-pixel padding around this region, and crop the original image to center the abnormality;
(ii) \textbf{ellipse.} We fit an optimal ellipse to the largest contour extracted from the lesion mask. 
(iii) \textbf{contour.} We smooth the lesion mask using Gaussian blur, detect its contours, and refine them with polygon approximation.
(iv) \textbf{bounding box.} We identify the minimum bounding box that completely contains the annotated lesion region, and add a 10-pixel padding around.

To simulate radiologists approaching CT images with prior knowledge, we provide GPT-4o with patient information (complaints, medical history, etc) to generate text queries that inquire specific abnormality based on such preliminary information. We use the following prompt template:
{
    \begin{tcolorbox}
Assuming a CT image has one or more abnormal findings, I will provide detailed information about them. Please help me generate prompts to test a VLM's ability to localize and analyze specific abnormalities.\\
\\
Requirements:\\
1. Generate prompts in English.\\
2. Ensure accurate information. The prompt content must stem from the real abnormality information, CT image reports, and doctor's discussion results I provided. You may use equivalents expressions in medical terminology, but do not introduce any content beyond these provided information.\\
3. Clear indication. Ensure each prompt refers to the corresponding abnormality without confusion with other abnormalities in the image.\\
4. Clear task. Each prompt should end with a clear request for the VLM to perform localization and analysis tasks.\\
5. Simulate realistic pre-examination clinical queries. Create prompts that reflect how clinicians approach CT images with preliminary information (such as the patient's complaints, medical history, other test results, etc., if available) and medical knowledge (such as common abnormalities associated with the patient's information or typically found in this imaged region), without being overly specific. Importantly, prompts should be broad enough to guide examination of suspicious areas and must not include detailed descriptions or conclusions from findings, reports and discussion results that would only be available after examining the CT image, such as specific abnormality details, exact measurements, precise locations, or definitive characteristics that could only be determined after image interpretation.\\
Some appropriate examples:\\
\$Some Example Queries\$\\
6. Avoid data leakage. To evaluate the VLM's ability to localize and analyze abnormalities independently, do not provide complete findings or diagnostic conclusions in the query. This prevents the VLM from bypassing the analytical process by retrieving answers directly from the prompt, while maintaining challenge authenticity.\\
7. Diversity. Generate at least 1 and at most 5 prompts with different perspectives for each abnormality. Each prompt should have clearly different focuses, avoiding content redundancy.\\

The information about all the abnormalities on the CT images:\\
\$Abnormal Findings\$

The clinical presentation of the patient corresponding to this CT image is:\\
\$Presentation\$

The overall report for the CT image containing these abnormalities is:\\
\$Whole Report\$

The doctors' discussion results for the patient corresponding to this CT image are:\\
\$Impression\$

Your response must follow this exact format:\\
\$Output Template\$
    \end{tcolorbox}
}

\clearpage

\subsection{Detailed Distribution of \dataset~}
~\label{sec:detailed_distribution}

\begin{table*}[h]
\renewcommand{\arraystretch}{1.3} % 增加行间距
\small
\center
% \setlength\tabcolsep{31pt}
% \resizebox{0.8\textwidth}{!}{
\begin{tabular}{lclccc}
\toprule
\rowcolor{lightgray} {\textbf{Organ}} & \textbf{Anatomical Structure} & \textbf{Category} & \textbf{Axial} & \textbf{Coronal} & \textbf{Sagittal} \\ \midrule

\multirow{33}{*}{Brain} & \multirow{14}{*}{Cerebral parenchyma} & Brain parenchymal atrophy & 31 & 2 & 4 \\
& & Brain parenchymal edema & 49 & 13 & 4 \\
& & Brain parenchymal soft tissue mass & 280 & 57 & 57 \\
& & \cellcolor[gray]{0.95}Brain parenchymal & \multirow{2}{*}{52} & \multirow{2}{*}{1} & \multirow{2}{*}{3} \\
& & \cellcolor[gray]{0.95}thin-walled cystic mass & & & \\
& & \cellcolor[gray]{0.9}Brain parenchymal & \multirow{2}{*}{17} & \multirow{2}{*}{0} & \multirow{2}{*}{1} \\
& & \cellcolor[gray]{0.9}thick-walled cystic mass & & & \\
& & \cellcolor[gray]{0.95}Brain parenchymal & \multirow{2}{*}{198} & \multirow{2}{*}{42} & \multirow{2}{*}{31} \\
& & \cellcolor[gray]{0.95}hemorrhage or contusion & & & \\
& & Acute infarct & 88 & 1 & 0 \\
& &  Lacunar infarct  & 54 & 11 & 8 \\
& &  Encephalomalacia & 27 & 6 & 3 \\
& & \cellcolor[gray]{0.95}Brain parenchymal & \multirow{2}{*}{102} & \multirow{2}{*}{31} & \multirow{2}{*}{23} \\
& & \cellcolor[gray]{0.95}morphological altrenation & & & \\
& & Other non-mass effect lesions & 44 & 5 & 5 \\
& & Intracranial air & 14 & 8 & 10 \\
& & Hyperdense lesions in brain parenchyma & 72 & 24 & 44 \\
& & Others & 35 & 13 & 6 \\
\cline{2-6}
& \multirow{5}{*}{Ventricles and cisterns} & Ventricular or cisternal enlargement & 122 & 25 & 25 \\
& &  Ventricular or cisternal soft tissue mass & 49 & 8 & 6 \\
& &  Ventricular or cisternal cystic mass  & 13 & 1 & 0 \\
& &  Ventricular or cisternal hemorrhage & 60 & 4 & 3 \\
& & Others & 11 & 2 & 3 \\
\cline{2-6}
& \multirow{6}{*}{\makecell{Meninges (including dura mater, \\ pia mater and arachnoid mater)}} & Meningeal cystic mass & 4 & 0 & 0 \\
& & Meningeal soft tissue mass & 8 & 9 & 6 \\
& & Meningeal hemorrhage  & 137 & 27 & 10 \\
& &  Meningeal effusion & 32 & 9 & 3 \\
& & Meningeal thickening & 6 & 0 & 0 \\
& & Others & 4 & 0 & 2 \\
\cline{2-6}
& \multirow{8}{*}{Pituitary and Sellar Region} &  Pituitary stalk thickening  & 0 & 0 & 0 \\
& & Pituitary stalk lateral displacement & 0 & 0 & 0 \\
& & Pituitary enlargement & 0 & 0 & 2 \\
& & Pituitary atrophy & 0 & 2 & 4 \\
& & Pituitary calcification & 0 & 0 & 3 \\
& & Pituitary or sella region soft tissue mass & 20 & 16 & 27 \\
& &  Pituitary or sella region cysitc mass  & 3 & 5 & 6 \\
& & Others & 1 & 0 & 0 \\

\bottomrule
\end{tabular}
% }
\caption{Detailed distribution of abnormality categories in \dataset~.}
\label{tab:detailed_distribution1}
\end{table*}

\begin{table*}[h]
\renewcommand{\arraystretch}{1.3} % 增加行间距
\center
\small
% \resizebox{0.8\textwidth}{!}{
\begin{tabular}{lclccc}
\toprule
\rowcolor{lightgray} {\textbf{Organ}} & \textbf{Anatomical Structure} & \textbf{Category} & \textbf{Axial} & \textbf{Coronal} & \textbf{Sagittal} \\ \midrule

\multirow{18}{*}{Spine} & \multirow{6}{*}{Spinal Cord} & Spinal cord compression & 3 & 0 & 2 \\
& & Spinal cord soft tissue mass & 12 & 1 & 6 \\
& & Spinal cord hemorrhage or contusion & 0 & 0 & 0 \\
& & Spinal morphological alteration & 0 & 0 & 0 \\
& & Syringomyelia & 0 & 0 & 0 \\
& & Others & 4 & 0 & 1 \\
\cline{2-6}
& \multirow{6}{*}{Intervertebral disc} & \cellcolor[gray]{0.95}Intervertebral disc & \multirow{2}{*}{4} & \multirow{2}{*}{2} & \multirow{2}{*}{1} \\
& & \cellcolor[gray]{0.95}morphological alteration & & & \\
& & \cellcolor[gray]{0.9}Intervertebral disc & \multirow{2}{*}{0} & \multirow{2}{*}{0} & \multirow{2}{*}{0} \\
& & \cellcolor[gray]{0.9}ossification or calcification & & & \\
& & Intervertebral disc gas & 4 & 3 & 2 \\
& &  Intervertebral disc extrusion & 0 & 0 & 0 \\
& & Intervertebral disc soft tissue mass & 5 & 1 & 1 \\
& & Others & 0 & 0 & 1 \\
\cline{2-6}
& \multirow{6}{*}{\makecell{Spinal meninge (including dura mater, \\ arachnoid mater, and pia mater)}} &  Spinal meningeal hemorrhage & 0 & 0 & 0 \\
& &  Spinal meningeal effusion & 0 & 0 & 1 \\
& &  Spinal meningeal thickening & 0 & 0 & 0 \\
& & Meningeal soft tissue mass & 7 & 2 & 6 \\
& &  Meningeal cystic mass  & 0 & 2 & 0 \\
& & Others & 2 & 1 & 5 \\
\hline

\multirow{16}{*}{Eye} & \multirow{7}{*}{Eyeball} & Eyeball atrophy & 4 & 1 & 2 \\
& & Eyeball positioning or morphological alteration & 44 & 7 & 6 \\
& & Eyeball density alteration & 45 & 20 & 14 \\
& & Eyeball soft tissue mass & 19 & 3 & 1 \\
& & \cellcolor[gray]{0.95}Eyeball wall thickening & \multirow{2}{*}{5} & \multirow{2}{*}{3} & \multirow{2}{*}{0} \\
& & \cellcolor[gray]{0.95}morphological alteration & & & \\
& & Complete or partial absence of eyeball structure & 2 & 0 & 0 \\
& & Others & 2 & 2 & 0 \\
\cline{2-6}
& \multirow{7}{*}{Ocular Adnexa} & Orbital density changes & 54 & 36 & 15 \\
& & \cellcolor[gray]{0.95}Intraorbital gas & \multirow{2}{*}{4} & \multirow{2}{*}{4} & \multirow{2}{*}{3} \\
& & \cellcolor[gray]{0.95}(e.g., soft tissue mass, fluid accumulation) & & & \\
& & Extraocular muscle hypertrophy or atrophy & 16 & 43 & 11 \\
& & Optic nerve thickening or soft tissue mass & 7 & 3 & 7 \\
& & Optic nerve atrophy & 0 & 0 & 0 \\
& & Others & 3 & 2 & 3 \\
\cline{2-6}
& \multirow{5}{*}{\makecell{Lacrimal gland \\and lacrimal sac}} & \cellcolor[gray]{0.95}Lacrimal gland and lacrimal sac & \multirow{2}{*}{19} & \multirow{2}{*}{12} & \multirow{2}{*}{5} \\
& & \cellcolor[gray]{0.95}enlargement or mass & & & \\
& & \cellcolor[gray]{0.9}Lacrimal gland and lacrimal sac & \multirow{2}{*}{4} & \multirow{2}{*}{2} & \multirow{2}{*}{0} \\
& & \cellcolor[gray]{0.9}calcification or fluid & & & \\
& & Others & 1 & 0 & 1 \\
\hline

\multirow{3}{*}{Ear} & \multirow{3}{*}{External Ear} & External auditory canal stenosis or atresia & 5 & 1 & 2 \\
& & External auditory canal soft tissue mass & 11 & 5 & 1 \\
& & Others & 1 & 1 & 0 \\

\bottomrule
\end{tabular}
\caption{(Continued) Detailed distribution of abnormality categories in \dataset~.}
% }
\label{tab:detailed_distribution2}
\end{table*}

\begin{table*}[h]
\renewcommand{\arraystretch}{1.3} % 增加行间距
\center
\small
% \resizebox{0.8\textwidth}{!}{
\begin{tabular}{lclccc}
\toprule
\rowcolor{lightgray} {\textbf{Organ}} & \textbf{Anatomical Structure} & \textbf{Category} & \textbf{Axial} & \textbf{Coronal} & \textbf{Sagittal} \\ \midrule

\multirow{11}{*}{Ear} & \multirow{6}{*}{Middle Ear} & Middle ear fluid or hemorrhage & 5 & 0 & 0 \\
& & Middle ear soft tissue mass & 23 & 23 & 3 \\
& & Ossicular chain destruction or deformity & 7 & 2 & 0 \\
& & Tympanic membrane thickening or calcification & 2 & 2 & 0 \\
& & Middle ear gas density change & 6 & 0 & 1 \\
& & Others & 1 & 2 & 0 \\
\cline{2-6}
& \multirow{5}{*}{Inner Ear} & Inner ear congenital structural alteration & 2 & 1 & 0 \\
& & Inner ear bone destruction or sclerosis & 11 & 4 & 0 \\
& & Internal auditory canal enlargement or narrowing & 2 & 0 & 0 \\
& & Labyrinthine structural alteration & 4 & 2 & 0 \\
& & Others & 1 & 1 & 0 \\
\hline
\multirow{12}{*}{Sinus} & \multirow{6}{*}{Sinus cavity} & Sinus effusion & 18 & 8 & 0 \\
& & Sinus hemorrhage & 6 & 3 & 3 \\
& & Sinus soft tissue mass & 98 & 85 & 33 \\
& & Sinus cystic mass & 8 & 5 & 1 \\
& & Sinus mucosal thickening & 24 & 25 & 7 \\
& & Others & 9 & 3 & 2 \\
\cline{2-6}
& \multirow{3}{*}{Sinus ostium} & Sinus obstruction or stenosis & 1 & 4 & 0 \\
& & Sinus widening & 0 & 0 & 1 \\
& & Others & 1 & 1 & 0 \\
\cline{2-6}
& \multirow{3}{*}{Nasal septum} & Nasal septum deviation or thickening & 7 & 6 & 0 \\
& & Nasal septum perforation or defect & 1 & 1 & 0 \\
& & Others & 6 & 2 & 0 \\
\hline

\multirow{17}{*}{Pharynx} & \multirow{6}{*}{\makecell{Pharynx (including nasopharynx, \\ oropharynx, and hypopharynx)}} & Pharyngeal narrowing or obstruction & 4 & 2 & 1 \\
& & Pharyngeal soft tissue mass & 46 & 14 & 10 \\
& & Pharyngeal cystic mass & 10 & 2 & 4 \\
& & Pharyngeal wall thickening & 4 & 3 & 0 \\
& & Pharyngeal foreign body & 0 & 0 & 0 \\
& & Others & 3 & 4 & 1 \\
\cline{2-6}
& \multirow{7}{*}{Larynx} & Laryngeal narrowing or obstruction & 3 & 2 & 0 \\
& & Vocal cord asymmetry & 2 & 0 & 0 \\
& & Vocal cord soft tissue mass & 6 & 1 & 1 \\
& & Laryngeal cartilage calcification & 0 & 0 & 0 \\
& & Laryngeal cystic mass & 1 & 1 & 0 \\
& & Laryngeal foreign body & 1 & 0 & 0 \\
& & Others & 1 & 0 & 0 \\
\cline{2-6}
& \multirow{4}{*}{Pharyngeal space} & Pharyngeal space soft tissue mass & 46 & 23 & 16 \\
& & Pharyngeal space cystic mass & 13 & 6 & 2 \\
& & Pharyngeal emphysema & 7 & 0 & 0 \\
& & Others & 5 & 7 & 1 \\

\bottomrule
\end{tabular}
% }
\caption{(Continued) Detailed distribution of abnormality categories in \dataset~.}
\label{tab:detailed_distribution3}
\end{table*}

\begin{table*}[h]
\renewcommand{\arraystretch}{1.3} % 增加行间距
\center
\small
% \resizebox{.75\textwidth}{!}{
\begin{tabular}{lclccc}
\toprule
\rowcolor{lightgray} {\textbf{Organ}} & \textbf{Anatomical Structure} & \textbf{Category} & \textbf{Axial} & \textbf{Coronal} & \textbf{Sagittal} \\ \midrule

\multirow{4}{*}{Pharynx} & \multirow{4}{*}{Pharyngeal space} & Pharyngeal space soft tissue mass & 46 & 23 & 16 \\
& & Pharyngeal space cystic mass & 13 & 6 & 2 \\
& & Pharyngeal emphysema & 7 & 0 & 0 \\
& & Others & 5 & 7 & 1 \\
\hline

\multirow{6}{*}{Parotid gland} & \multirow{6}{*}{Parotid gland} & Parotid gland enlargement & 20 & 2 & 0 \\
& & Parotid gland atrophy & 4 & 0 & 0 \\
& & Parotid gland soft tissue mass & 33 & 2 & 2 \\
& & Parotid gland cystic mass & 16 & 0 & 0 \\
& & Parotid gland calcification or stone & 15 & 0 & 2 \\
& & Others & 10 & 0 & 0 \\
\hline

\multirow{7}{*}{Thyroid gland} & \multirow{7}{*}{Thyroid gland} & Thyroid enlargement & 12 & 2 & 2 \\
& & Thyroid atrophy & 1 & 0 & 0 \\
& & Thyroid soft tissue mass & 41 & 4 & 0 \\
& & Thyroid cystic mass & 3 & 1 & 0 \\
& & Thyroid calcification & 1 & 0 & 0 \\
& & Ectopic thyroid gland & 5 & 1 & 0 \\
& & Others & 6 & 0 & 1 \\
\hline

\multirow{7}{*}{Trachea} & \multirow{7}{*}{Tracheal lumen} & Tracheal stenosis or obstruction & 12 & 4 & 7 \\
& & Tracheal dilatation & 11 & 5 & 2 \\
& & Tracheal soft tissue mass & 12 & 6 & 1 \\
& & Tracheal wall thickening & 9 & 0 & 0 \\
& & Tracheal wall calcification & 3 & 1 & 4 \\
& & Tracheal wall defect & 3 & 0 & 2 \\
& & Others & 1 & 0 & 2 \\
\hline

\multirow{16}{*}{Lung} & \multirow{12}{*}{pulmonary parenchyma} & Atelectasis & 106 & 10 & 3 \\
& & Incomplete lung expansion & 52 & 3 & 1 \\
& & Pulmonary consolidation & 203 & 24 & 2 \\
& & Pulmonary ground-glass opacities & 343 & 52 & 5 \\
& & Pulmonary emphysema & 45 & 10 & 2 \\
& & Pulmonary solitary nodule or mass & 226 & 46 & 10 \\
& & Pulmonary diffusely & \multirow{2}{*}{285} & \multirow{2}{*}{76} & \multirow{2}{*}{3} \\
& & distributed multiple nodules & & & \\
& & Pulmonary parenchymal fibrosis & 42 & 12 & 0 \\
& & Pulmonary thin-walled cavitation & 48 & 6 & 1 \\
& & Pulmonary thick-walled cavities & 44 & 1 & 0 \\
& & Pulmonary cystic mass & 53 & 14 & 1 \\
& & Others & 14 & 6 & 1 \\
\cline{2-6}
& \multirow{4}{*}{Bronchi} & Bronchiectasis & 97 & 19 & 6 \\
& & Bronchial wall thickening, & \multirow{2}{*}{15} & \multirow{2}{*}{11} & \multirow{2}{*}{5} \\
& & stenosis, or occlusion & & & \\
& & Bronchial foreign body & 4 & 1 & 0 \\
& & Others & 3 & 0 & 1 \\

\bottomrule
\end{tabular}
% }
\caption{(Continued) Detailed distribution of abnormality categories in \dataset~.}
\label{tab:detailed_distribution4}
\end{table*}

\begin{table*}[h]
\renewcommand{\arraystretch}{1.3} % 增加行间距
\center
\small
% \resizebox{0.8\textwidth}{!}{
\begin{tabular}{lclccc}
\toprule
\rowcolor{lightgray} {\textbf{Organ}} & \textbf{Anatomical Structure} & \textbf{Category} & \textbf{Axial} & \textbf{Coronal} & \textbf{Sagittal} \\ \midrule

\multirow{9}{*}{Lung} & \multirow{3}{*}{Lung interstitial} & Lung interstitial fibrosis and thickening & 78 & 10 & 0 \\
& & Honeycomb lung & 34 & 0 & 0 \\
& & Others & 2 & 0 & 0 \\
\cline{2-6}
& \multirow{6}{*}{Pleura} & Pleural thickening & 39 & 10 & 1 \\
& & Pleural effusion & 268 & 27 & 3 \\
& & Pneumothorax & 36 & 2 & 0 \\
& & Pleural calcification & 24 & 0 & 1 \\
& & Pleural soft tissue mass & 50 & 8 & 0 \\
& & Others & 8 & 1 & 1 \\
\hline

\multirow{6}{*}{Mediastinum} & \multirow{6}{*}{Mediastinal soft tissue} & Mediastinal shift & 29 & 4 & 0 \\
& & Mediastinal soft tissue mass & 200 & 31 & 15 \\
& & Mediastinal cystic mass & 19 & 0 & 2 \\
& & Mediastinal hemorrhage & 15 & 0 & 0 \\
& & Mediastinal emphysema & 58 & 7 & 1 \\
& & Others & 10 & 0 & 0 \\
\hline

\multirow{4}{*}{Diaphragm} & \multirow{4}{*}{Diaphragm} & Diaphragmatic hernia & 19 & 16 & 12 \\
& & Diaphragmatic elevation & 5 & 4 & 0 \\
& & Diaphragmatic soft tissue mass & 7 & 2 & 0 \\
& & Others & 0 & 2 & 0 \\
\hline

\multirow{6}{*}{Thymus} & \multirow{6}{*}{Thymic parenchyma} & Thymic enlargement & 1 & 0 & 0 \\
& & Thymic atrophy & 0 & 0 & 0 \\
& & Thymic soft tissue mass & 6 & 0 & 0 \\
& & Thymic cystic mass & 0 & 0 & 0 \\
& & Thymic calcification & 4 & 0 & 0 \\
& & Others & 1 & 0 & 0 \\
\hline

\multirow{11}{*}{Oral cavity} & \multirow{4}{*}{Oral Soft Tissue} & Oral soft tissue mass & 8 & 0 & 0 \\
& & Oral cystic mass & 3 & 0 & 0 \\
& & Oral soft tissue calcification & 0 & 0 & 0 \\
& & Others & 0 & 0 & 0 \\
\cline{2-6}
& \multirow{7}{*}{Teeth and alveolar bone} & Dental developmental anomalies & 1 & 0 & 0 \\
& & Dental positional anomalies & 3 & 4 & 5 \\
& & Dental calcification or caries & 3 & 0 & 0 \\
& & Alveolar bone resorption or hyperplasia & 3 & 0 & 0 \\
& & Alveolar soft tissue mass & 8 & 2 & 0 \\
& & Alveolar cystic mass & 8 & 1 & 1 \\
& & Others & 0 & 1 & 0 \\
\hline

\multirow{3}{*}{Heart} & \multirow{3}{*}{\makecell{Cardiac chambers \\(atrium or ventricle)}} & Cardiac chamber enlargement & 41 & 7 & 3 \\
& & Cardiac chamber mass & 28 & 7 & 2 \\
& & Others & 17 & 1 & 0 \\

\bottomrule
\end{tabular}
% }
\caption{(Continued) Detailed distribution of abnormality categories in \dataset~.}
\label{tab:detailed_distribution5}
\end{table*}

\begin{table*}[h]
\renewcommand{\arraystretch}{1.3} % 增加行间距
\center
\small
% \resizebox{.75\textwidth}{!}{
\begin{tabular}{lclccc}
\toprule
\rowcolor{lightgray} {\textbf{Organ}} & \textbf{Anatomical Structure} & \textbf{Category} & \textbf{Axial} & \textbf{Coronal} & \textbf{Sagittal} \\ \midrule

\multirow{16}{*}{Heart} & \multirow{5}{*}{Myocardium} & Myocardial hypertrophy & 4 & 1 & 1 \\
& & Myocardial thinning & 3 & 0 & 0 \\
& & Myocardial calcification & 4 & 1 & 0 \\
& & Myocardial density alteration & 6 & 0 & 0 \\
& & Others & 3 & 3 & 0 \\
\cline{2-6}
& \multirow{2}{*}{Heart valve} & Valvular calcification & 3 & 0 & 0 \\
& & Others & 3 & 0 & 0 \\
\cline{2-6}
& \multirow{6}{*}{Pericardium} & Pericardial thickening & 9 & 0 & 2 \\
& & Pericardial calcification & 6 & 1 & 0 \\
& & Pericardial effusion & 54 & 7 & 2 \\
& & Pericardial hemorrhage & 2 & 0 & 0 \\
& & Pericardial emphysema & 6 & 1 & 0 \\
& & Others & 4 & 0 & 0 \\
\cline{2-6}
& \multirow{3}{*}{Coronary arteries} & Coronary artery myocardial bridge & 1 & 0 & 0 \\
& & Coronary artery dilation & 0 & 0 & 0 \\
& & Others & 3 & 1 & 0 \\
\hline

\multirow{24}{*}{\makecell{Vascular \\structure}} & \multirow{12}{*}{Artery} & Arterial widening & 66 & 24 & 9 \\
& & Aneurysm & 138 & 44 & 39 \\
& & \cellcolor[gray]{0.95}Atherosclerotic plaque & \multirow{2}{*}{39} & \multirow{2}{*}{13} & \multirow{2}{*}{4} \\
& & \cellcolor[gray]{0.95}of arterial wall & & & \\
& & Arterial wall ulcer & 8 & 2 & 2 \\
& & \cellcolor[gray]{0.95}Arterial dissection or & \multirow{2}{*}{22} & \multirow{2}{*}{15} & \multirow{2}{*}{8} \\
& & \cellcolor[gray]{0.95}intramural hematoma & & & \\
& & Arterial stenosis & 24 & 8 & 10 \\
& & Arterial occlusion & 55 & 9 & 0 \\
& & Arterial filling defect & 80 & 23 & 8 \\
& & Arterial contour abnormality & 51 & 23 & 23 \\
& & Arteriovenous fistula & 16 & 7 & 1 \\
& & Arterial wall inflammatory exudate & 4 & 0 & 0 \\
& & Others & 39 & 9 & 7 \\
\cline{2-6}
& \multirow{9}{*}{Vein} & Venous dilation & 41 & 4 & 6 \\
& & Varicosity & 23 & 11 & 9 \\
& & Venous wall inflammation & 0 & 0 & 0 \\
& & Venous stenosis & 7 & 1 & 1 \\
& & Venous occlusion & 5 & 2 & 1 \\
& & Venous filling defect & 48 & 12 & 3 \\
& & Venous morphological abnormality & 29 & 6 & 3 \\
& & Venous wall inflammatory exudate & 2 & 1 & 0 \\
& & Others & 13 & 8 & 0 \\
\cline{2-6}
& \multirow{3}{*}{Capillary} & Capillary dilation & 0 & 0 & 0 \\
& & Capillary malformation proliferation & 1 & 0 & 0 \\
& & Others & 0 & 0 & 0 \\

\bottomrule
\end{tabular}
% }
\caption{(Continued) Detailed distribution of abnormality categories in \dataset~.}
\label{tab:detailed_distribution6}
\end{table*}

\begin{table*}[h]
\renewcommand{\arraystretch}{1.3} % 增加行间距
\center
\small
% \resizebox{.75\textwidth}{!}{
\begin{tabular}{lclccc}
\toprule
\rowcolor{lightgray} {\textbf{Organ}} & \textbf{Anatomical Structure} & \textbf{Category} & \textbf{Axial} & \textbf{Coronal} & \textbf{Sagittal} \\ \midrule

\multirow{22}{*}{\makecell{Gastrointestinal \\tract \\(including \\esophagus, \\stomach, \\and intestines)}} & \multirow{7}{*}{Gastrointestinal lumen} & Gastrointestinal dilatation & 265 & 212 & 112 \\
& & Gastrointestinal narrowing & 20 & 5 & 6 \\
& & Gastrointestinal foreign body & 2 & 6 & 0 \\
& & Gastrointestinal air-fluid level & 31 & 8 & 5 \\
& & \cellcolor[gray]{0.95}Gastrointestinal luminal & \multirow{2}{*}{17} & \multirow{2}{*}{6} & \multirow{2}{*}{4} \\
& & \cellcolor[gray]{0.95}contents abnormal density & & & \\
& & Others & 8 & 3 & 1 \\
\cline{2-6}
& \multirow{7}{*}{Gastrointestinal tract wall} & Gastrointestinal wall thickening & 398 & 212 & 120 \\
& & Gastrointestinal wall mass & 103 & 56 & 26 \\
& & \cellcolor[gray]{0.95}Gastrointestinal wall & \multirow{2}{*}{43} & \multirow{2}{*}{12} & \multirow{2}{*}{4} \\
& & \cellcolor[gray]{0.95}rupture and perforation & & & \\
& & Gastrointestinal wall ulceration & 0 & 0 & 0 \\
& & Others & 7 & 9 & 5 \\
\cline{2-6}
& \multirow{3}{*}{\makecell{Gastrointestinal \\positioning \\abnormality}} & \cellcolor[gray]{0.95}Gastrointestinal herniation & \multirow{2}{*}{117} & \multirow{2}{*}{67} & \multirow{2}{*}{30} \\
& & \cellcolor[gray]{0.95}or deformity & & & \\
& & Others & 5 & 1 & 1 \\
\cline{2-6}
& \multirow{6}{*}{\makecell{Gastrointestinal \\morphological \\ abnormalities}} & Gastrointestinal diverticulum & 53 & 35 & 15 \\
& & Gastrointestinal malrotation and volvulus & 17 & 17 & 3 \\
& & \cellcolor[gray]{0.95}Gastrointestinal annular or & \multirow{2}{*}{28} & \multirow{2}{*}{22} & \multirow{2}{*}{10} \\
& & \cellcolor[gray]{0.95}concentric abnormality & & & \\
& & Others & 2 & 1 & 2 \\
\cline{2-6}
& \multirow{5}{*}{Mesentery} & Mesenteric volvulus & 3 & 4 & 2 \\
& & Mesenteric edema & 10 & 3 & 3 \\
& & Mesenteric panniculitis & 26 & 6 & 6 \\
& & Mesenteric soft tissue mass & 28 & 6 & 9 \\
& & Others & 0 & 4 & 2 \\
\hline

\multirow{11}{*}{Liver} & \multirow{11}{*}{Hepatic parenchyma} & Hepatic parenchymal morphological alteration & 99 & 34 & 11 \\
& & Hepatic parenchymal hyperdensity & 18 & 2 & 0 \\
& & Hepatic parenchymal hypodensity & 71 & 25 & 9 \\
& & Hepatic parenchymal soft tissue mass & 439 & 58 & 20 \\
& & Hepatic parenchymal thin-walled cystic mass & 124 & 23 & 10 \\
& & Hepatic parenchymal thick-walled cystic mass & 21 & 6 & 4 \\
& & Liver contusion or hemorrhage & 10 & 3 & 0 \\
& & Intrahepatic bile duct emphysema & 13 & 1 & 0 \\
& & Intrahepatic bile duct fluid accumulation & 0 & 0 & 0 \\
& & Intrahepatic biliary stones or calcification & 7 & 0 & 0 \\
& & Others & 16 & 4 & 1 \\
\hline

\multirow{5}{*}{Gallbladder} & \multirow{5}{*}{Gallbladder lumen} & Gallbladder distension & 39 & 10 & 1 \\
& & Gallbladder atrophy or shrinkage & 4 & 1 & 0 \\
& & Gallbladder stone & 87 & 22 & 2 \\
& & Gallbladder contents density change & 32 & 5 & 0 \\
& & Others & 3 & 1 & 0 \\

\bottomrule
\end{tabular}
% }
\caption{(Continued) Detailed distribution of abnormality categories in \dataset~.}
\label{tab:detailed_distribution7}
\end{table*}

\begin{table*}[h]
\renewcommand{\arraystretch}{1.3} % 增加行间距
\center
\small
% \resizebox{.75\textwidth}{!}{
\begin{tabular}{lclccc}
\toprule
\rowcolor{lightgray} {\textbf{Organ}} & \textbf{Anatomical Structure} & \textbf{Category} & \textbf{Axial} & \textbf{Coronal} & \textbf{Sagittal} \\ \midrule

\multirow{17}{*}{Gallbladder} & \multirow{5}{*}{Gallbladder wall} & Gallbladder wall thickening & 53 & 11 & 1 \\
& & Gallbladder wall calcification & 9 & 1 & 0 \\
& & Gallbladder wall mass & 16 & 2 & 1 \\
& & Gallbladder wall rupture & 11 & 1 & 0 \\
& & Others & 3 & 1 & 1 \\
\cline{2-6}
& \multirow{3}{*}{Gallbladder morphology} & Gallbladder position alteration & 1 & 0 & 0 \\
& & Gallbladder congenital morphological variation & 0 & 0 & 0 \\
& & Others & 1 & 1 & 0 \\
\cline{2-6}
& \multirow{9}{*}{Extrahepatic bile ducts} & Extrahepatic bile duct dilation & 26 & 8 & 0 \\
& & Extrahepatic bile duct wall thickening & 2 & 0 & 0 \\
& & Extrahepatic bile duct soft tissue mass & 1 & 0 & 1 \\
& & Extrahepatic bile duct cystic mass & 1 & 1 & 0 \\
& & Extrahepatic bile duct stenosis or obstruction & 2 & 0 & 0 \\
& & Extrahepatic bile duct content density alteration & 6 & 1 & 0 \\
& & Extrahepatic bile duct stone & 15 & 5 & 0 \\
& & Extrahepatic bile duct injury or rupture & 0 & 0 & 0 \\
& & Others & 1 & 0 & 0 \\
\hline

\multirow{12}{*}{Pancreas} & \multirow{6}{*}{Pancreatic parenchyma} & Pancreatic parenchymal soft tissue mass & 137 & 8 & 4 \\
& & Pancreatic cystic mass & 69 & 7 & 4 \\
& & Pancreatic calcification & 20 & 1 & 0 \\
& & Pancreatic enlargement & 60 & 1 & 1 \\
& & Pancreatic atrophy & 8 & 1 & 1 \\
& & Others & 16 & 0 & 0 \\
\cline{2-6}
& \multirow{3}{*}{Pancreatic duct} & Pancreatic ductal dilatation & 13 & 3 & 0 \\
& & Pancreatic ductal stone & 3 & 0 & 0 \\
& & Others & 0 & 0 & 0 \\
\cline{2-6}
& \multirow{3}{*}{Pancreatic morphology} & Pancreatic congenital anomaly & 5 & 0 & 0 \\
& & Pancreatic positional displacement & 1 & 0 & 0 \\
& & Others & 1 & 0 & 0 \\
\hline

\multirow{8}{*}{Spleen} & \multirow{6}{*}{Splenic parenchyma} & Splenic parenchymal calcification & 15 & 0 & 1 \\
& & Splenic parenchymal soft tissue mass & 52 & 11 & 3 \\
& & Splenic parenchymal cystic mass & 27 & 10 & 0 \\
& & Splenic parenchymal infarct & 19 & 1 & 0 \\
& & Splenic parenchymal rupture & 12 & 1 & 0 \\
& & Others & 18 & 2 & 0 \\
\cline{2-6}
& \multirow{2}{*}{Spleen morphology} & Spleen enlargement & 102 & 22 & 5 \\
& & Others & 17 & 1 & 0 \\

\bottomrule
\end{tabular}
% }
\caption{(Continued) Detailed distribution of abnormality categories in \dataset~.}
\label{tab:detailed_distribution8}
\end{table*}

\begin{table*}[h]
\renewcommand{\arraystretch}{1.3} % 增加行间距
\center
\small
% \resizebox{.75\textwidth}{!}{
\begin{tabular}{lclccc}
\toprule
\rowcolor{lightgray} {\textbf{Organ}} & \textbf{Anatomical Structure} & \textbf{Category} & \textbf{Axial} & \textbf{Coronal} & \textbf{Sagittal} \\ \midrule

\multirow{14}{*}{\makecell{Abdominopelvic \\peritoneum}} & \multirow{14}{*}{\makecell{Abdominopelvic \\peritoneum}} & Peritoneal inflammatory exudate & 75 & 19 & 10 \\
& & Peritoneal thickening & 19 & 3 & 1 \\
& & Peritoneal calcification & 5 & 1 & 0 \\
& & Peritoneal soft tissue mass & 364 & 135 & 90 \\
& & Peritoneal cystic mass & 102 & 34 & 22 \\
& & Abdominopelvic fluid & 280 & 96 & 43 \\
& & Abdominopelvic hemorrhage & 68 & 3 & 1 \\
& & Abdominopelvic free air & 61 & 17 & 10 \\
& & Retroperitoneal fibrosis & 1 & 0 & 0 \\
& & \cellcolor[gray]{0.95}Peritoneal or retroperitoneal & \multirow{2}{*}{17} & \multirow{2}{*}{18} & \multirow{2}{*}{4} \\
& & \cellcolor[gray]{0.95}lymph enlargement & & & \\
& & Extravasation of gastrointestinal content & 0 & 1 & 0 \\
& & Abdominopelvic contrast agent leakage & 22 & 13 & 4 \\
& & Others & 12 & 12 & 2 \\
\hline

\multirow{20}{*}{Kidney} & \multirow{7}{*}{Renal Parenchyma} & Renal parenchymal soft tissue mass & 150 & 81 & 29 \\
& & Renal parenchymal cystic mass & 239 & 68 & 21 \\
& & Renal parenchymal calcification & 9 & 9 & 2 \\
& & \cellcolor[gray]{0.95}Renal parenchymal or & \multirow{2}{*}{11} & \multirow{2}{*}{3} & \multirow{2}{*}{2} \\
& & \cellcolor[gray]{0.95}subcapsular hemorrhage & & & \\
& & Renal infarct & 8 & 3 & 0 \\
& & Others & 27 & 8 & 1 \\
\cline{2-6}
& \multirow{9}{*}{\makecell{Renal pelvis \\and ureter}} & Hydronephrosis & 81 & 63 & 3 \\
& & Ureteral dilatation & 35 & 29 & 4 \\
& & Ureteral stricture or obstruction & 1 & 1 & 0 \\
& & Ureteric stone & 59 & 60 & 19 \\
& & \cellcolor[gray]{0.95}Double renal pelvis and/or & \multirow{2}{*}{5} & \multirow{2}{*}{3} & \multirow{2}{*}{1} \\
& & \cellcolor[gray]{0.95}double ureter anomaly & & & \\
& & Renal pelvis soft tissue mass & 13 & 4 & 0 \\
& & Renal pelvic cystic mass & 1 & 3 & 0 \\
& & Others & 4 & 5 & 2 \\
\cline{2-6}
& \multirow{5}{*}{\makecell{Renal morphology and \\position abnormalities}} & Renal morphological anomaly & 33 & 20 & 1 \\
& & Renal enlargement & 56 & 16 & 1 \\
& & Renal atrophy & 23 & 6 & 2 \\
& & Ectopic or transplanted kidney & 17 & 13 & 5 \\
& & Others & 3 & 2 & 0 \\
\hline

\multirow{4}{*}{Bladder} & \multirow{4}{*}{Bladder cavity} & Bladder distention & 4 & 0 & 1 \\
& & Bladder stone & 24 & 5 & 3 \\
& & Bladder content density change & 13 & 2 & 1 \\
& & Others & 4 & 3 & 0 \\

\bottomrule
\end{tabular}
% }
\caption{(Continued) Detailed distribution of abnormality categories in \dataset~.}
\label{tab:detailed_distribution9}
\end{table*}

\begin{table*}[h]
\renewcommand{\arraystretch}{1.3} % 增加行间距
\center
\small
% \resizebox{.75\textwidth}{!}{
\begin{tabular}{lclccc}
\toprule
\rowcolor{lightgray} {\textbf{Organ}} & \textbf{Anatomical Structure} & \textbf{Category} & \textbf{Axial} & \textbf{Coronal} & \textbf{Sagittal} \\ \midrule

\multirow{15}{*}{Bladder} & \multirow{4}{*}{Bladder cavity} & Bladder distention & 4 & 0 & 1 \\
& & Bladder stone & 24 & 5 & 3 \\
& & Bladder content density change & 13 & 2 & 1 \\
& & Others & 4 & 3 & 0 \\
\cline{2-6}
& \multirow{8}{*}{Bladder wall} & Bladder wall diffuse thickening & 17 & 5 & 6 \\
& & Bladder wall calcification & 1 & 3 & 1 \\
& & \cellcolor[gray]{0.95}Bladder wall focal thickening & \multirow{2}{*}{69} & \multirow{2}{*}{7} & \multirow{2}{*}{3} \\
& & \cellcolor[gray]{0.95}or soft tissue mass & & & \\
& & Bladder wall defect or fistula & 4 & 0 & 1 \\
& & Bladder wall emphysema & 9 & 0 & 1 \\
& & Bladder wall diverticulum & 31 & 2 & 2 \\
& & Others & 1 & 0 & 0 \\
\cline{2-6}
& \multirow{3}{*}{Bladder morphology} & Bladder position displacement & 9 & 2 & 2 \\
& & Bladder morphological anomaly & 2 & 0 & 0 \\
& & Others & 1 & 0 & 2 \\
\hline

\multirow{6}{*}{Adrenal gland} & \multirow{6}{*}{Adrenal gland} & Adrenal soft tissue mass & 122 & 17 & 1 \\
& & Adrenal cystic mass & 32 & 0 & 0 \\
& & Adrenal calcification & 0 & 3 & 0 \\
& & Adrenal gland thickening & 16 & 0 & 0 \\
& & Adrenal atrophy & 0 & 0 & 0 \\
& & Others & 15 & 0 & 0 \\
\hline

\multirow{7}{*}{Prostate} & \multirow{7}{*}{Prostate} & Prostate enlargement & 18 & 12 & 14 \\
& & Prostate atrophy & 0 & 0 & 0 \\
& & Prostatic cystic mass & 2 & 0 & 0 \\
& & Prostatic soft tissue density anomaly & 2 & 0 & 1 \\
& & Prostatic calcification & 2 & 1 & 2 \\
& & Prostatic hemarrhage & 0 & 0 & 0 \\
& & Others & 0 & 0 & 0 \\
\hline

\multirow{4}{*}{Seminal vesicle} & \multirow{4}{*}{Seminal vesicle} & Seminal vesicle soft tissue mass & 1 & 0 & 0 \\
& & Seminal vesicle calcification & 0 & 0 & 1 \\
& & Seminal vesicle cystic mass & 3 & 0 & 0 \\
& & Others & 0 & 0 & 0 \\
\hline

\multirow{8}{*}{\makecell{Testes, epididymis, \\and scrotum}} & \multirow{8}{*}{Testis} & Testicular enlargement & 1 & 0 & 0 \\
& & Testicular atrophy & 0 & 0 & 0 \\
& & Testicular soft tissue mass & 1 & 0 & 0 \\
& & Testicular calcification & 0 & 0 & 0 \\
& & Testicular cystic mass & 0 & 0 & 0 \\
& & Testicular torsion & 0 & 0 & 0 \\
& & Testicular hemorrhage and rupture & 0 & 0 & 0 \\
& & Others & 0 & 0 & 0 \\

\bottomrule
\end{tabular}
% }
\caption{(Continued) Detailed distribution of abnormality categories in \dataset~.}
\label{tab:detailed_distribution10}
\end{table*}

\begin{table*}[h]
\renewcommand{\arraystretch}{1.3} % 增加行间距
\center
\small
% \resizebox{.75\textwidth}{!}{
\begin{tabular}{lclccc}
\toprule
\rowcolor{lightgray} {\textbf{Organ}} & \textbf{Anatomical Structure} & \textbf{Category} & \textbf{Axial} & \textbf{Coronal} & \textbf{Sagittal} \\ \midrule

\multirow{19}{*}{\makecell{Testes, epididymis, \\and scrotum}} & \multirow{8}{*}{Testis} & Testicular enlargement & 1 & 0 & 0 \\
& & Testicular atrophy & 0 & 0 & 0 \\
& & Testicular soft tissue mass & 1 & 0 & 0 \\
& & Testicular calcification & 0 & 0 & 0 \\
& & Testicular cystic mass & 0 & 0 & 0 \\
& & Testicular torsion & 0 & 0 & 0 \\
& & Testicular hemorrhage and rupture & 0 & 0 & 0 \\
& & Others & 0 & 0 & 0 \\
\cline{2-6}
& \multirow{5}{*}{Scrotum} & Scrotal effusion & 1 & 0 & 0 \\
& & Scrotal hematoma & 0 & 0 & 0 \\
& & Scrotal soft tissue mass & 1 & 0 & 0 \\
& & Scrotal wall thickening & 0 & 0 & 0 \\
& & Others & 2 & 0 & 0 \\
\cline{2-6}
& \multirow{6}{*}{Epididymis} & Epididymis enlargement & 0 & 0 & 0 \\
& & Epididymal soft tissue mass & 0 & 0 & 0 \\
& & Epididymal calcification & 0 & 0 & 0 \\
& & Epididymal cystic mass & 0 & 0 & 0 \\
& & Epididymal thickening & 0 & 0 & 0 \\
& & Others & 0 & 0 & 0 \\
\hline

\multirow{7}{*}{Penis} & \multirow{7}{*}{Penis} & Penile morphological anomaly & 0 & 0 & 0 \\
& & Penile soft tissue mass & 0 & 0 & 0 \\
& & Penile calcification & 0 & 0 & 0 \\
& & Urethral calculi or foreign body & 6 & 2 & 1 \\
& & Urethral Stricture & 0 & 0 & 0 \\
& & Urethral dilation & 1 & 0 & 0 \\
& & Others & 4 & 0 & 0 \\
\hline

\multirow{8}{*}{Uterus} & \multirow{8}{*}{Uterus} & Uterine morphological anomaly & 7 & 1 & 2 \\
& & Uterine enlargement & 15 & 3 & 6 \\
& & Uterine soft tissue mass & 50 & 4 & 7 \\
& & Uterine calcification & 7 & 0 & 2 \\
& & Uterine cavity effusion & 2 & 0 & 1 \\
& & Uterine cavity hemorrhage & 2 & 0 & 0 \\
& & Uterine cystic mass & 4 & 0 & 0 \\
& & Others & 4 & 1 & 1 \\
\hline

\multirow{7}{*}{Fallopian tube} & \multirow{7}{*}{Fallopian tube} & Fallopian tube thickening & 3 & 0 & 0 \\
& & Fallopian tube cystic mass & 2 & 0 & 1 \\
& & Fallopian tube soft tissue mass & 12 & 4 & 2 \\
& & Fallopian tube effusion & 2 & 0 & 0 \\
& & Fallopian tube hemorrhage & 0 & 0 & 0 \\
& & Fallopian tube calcification & 0 & 0 & 0 \\
& & Others & 0 & 0 & 0 \\

\bottomrule
\end{tabular}
% }
\caption{(Continued) Detailed distribution of abnormality categories in \dataset~.}
\label{tab:detailed_distribution10}
\end{table*}

\begin{table*}[h]
\renewcommand{\arraystretch}{1.3} % 增加行间距
\center
\small
% \resizebox{.75\textwidth}{!}{
\begin{tabular}{lclccc}
\toprule
\rowcolor{lightgray} {\textbf{Organ}} & \textbf{Anatomical Structure} & \textbf{Category} & \textbf{Axial} & \textbf{Coronal} & \textbf{Sagittal} \\ \midrule

\multirow{7}{*}{Ovary} & \multirow{7}{*}{Ovary} & Ovarian enlargement & 9 & 3 & 0 \\
& & Ovarian Atrophy & 0 & 0 & 0 \\
& & Ovarian cystic mass & 85 & 12 & 6 \\
& & Ovarian soft tissue mass & 96 & 6 & 3 \\
& & Ovarian calcification & 3 & 1 & 0 \\
& & Ovarian torsion & 9 & 0 & 0 \\
& & Others & 1 & 1 & 0 \\
\hline

\multirow{6}{*}{Vagina and vulva} & \multirow{6}{*}{Vagina and vulva} & Vaginal soft tissue mass & 0 & 0 & 0 \\
& & Vaginal cystic mass & 9 & 0 & 0 \\
& & Vaginal hemorrhage & 1 & 0 & 0 \\
& & Vaginal emphysema & 3 & 0 & 4 \\
& & Vaginal anatomical anomaly & 3 & 0 & 1 \\
& & Others & 6 & 0 & 0 \\
\hline

\multirow{14}{*}{Breast} & \multirow{6}{*}{Breast gland} & Breast gland enlargement & 2 & 0 & 0 \\
& & Breast gland atrophy & 0 & 0 & 0 \\
& & Breast gland soft tissue mass & 36 & 9 & 2 \\
& & Breast gland calcification & 0 & 0 & 0 \\
& & Breast gland cystic mass & 0 & 0 & 0 \\
& & Others & 6 & 0 & 0 \\
\cline{2-6}
& \multirow{3}{*}{Breast duct} & Breast duct dilation & 0 & 0 & 0 \\
& & Breast duct calcification & 0 & 0 & 0 \\
& & Others & 0 & 0 & 0 \\
\cline{2-6}
& \multirow{3}{*}{Nipple} & Nipple retraction & 0 & 0 & 0 \\
& & Nipple calcification & 0 & 0 & 0 \\
& & Others & 0 & 0 & 0 \\
\cline{2-6}
& \multirow{2}{*}{Areola} & Areola thickening & 1 & 0 & 0 \\
& & Others & 0 & 0 & 0 \\
\hline

\multirow{16}{*}{Skeletal system} & \multirow{16}{*}{Skeletal system} & Osteoporosis & 9 & 1 & 8 \\
& & Osteomalacia & 2 & 0 & 1 \\
& & Bone destruction or soft tissue mass & 219 & 124 & 117 \\
& & Bone cystic mass & 19 & 32 & 17 \\
& & Osseous sclerosis & 179 & 80 & 121 \\
& & Osteonecrosis & 64 & 13 & 24 \\
& & Bone fracture & 308 & 172 & 247 \\
& & Periosteal reaction & 23 & 9 & 5 \\
& & Periosteal thickening & 8 & 5 & 1 \\
& & Bone callus and post-fracture healing & 10 & 8 & 2 \\
& & Scar of fracture fixation removal & 0 & 0 & 0 \\
& & Bone deformation & 67 & 44 & 73 \\
& & Skeletal asymmetry & 10 & 2 & 1 \\
& & Cartilage calcification & 4 & 4 & 8 \\
& & Chondral calcification & 5 & 0 & 1 \\
& & Others & 30 & 50 & 26 \\

\bottomrule
\end{tabular}
% }
\caption{(Continued) Detailed distribution of abnormality categories in \dataset~.}
\label{tab:detailed_distribution10}
\end{table*}

\begin{table*}[h]
\renewcommand{\arraystretch}{1.3} % 增加行间距
\center
\small
% \resizebox{.75\textwidth}{!}{
\begin{tabular}{lclccc}
\toprule
\rowcolor{lightgray} {\textbf{Organ}} & \textbf{Anatomical Structure} & \textbf{Category} & \textbf{Axial} & \textbf{Coronal} & \textbf{Sagittal} \\ \midrule

\multirow{13}{*}{Joint} & \multirow{13}{*}{Joint} & Joint space narrowing & 6 & 12 & 3 \\
& & Joint space widening & 3 & 0 & 3 \\
& & Joint cartilage degradation & 1 & 0 & 2 \\
& & Joint cartilage calcification & 3 & 1 & 1 \\
& & Joint capsule thickening & 0 & 0 & 0 \\
& & Intra-articular effusion & 13 & 3 & 2 \\
& & Intra-articular hemorrhage & 4 & 0 & 0 \\
& & Intra-articular gas & 1 & 0 & 4 \\
& & \cellcolor[gray]{0.95}Joint periarticular & \multirow{2}{*}{2} & \multirow{2}{*}{0} & \multirow{2}{*}{0} \\
& & \cellcolor[gray]{0.95}soft tissue swelling & & & \\
& & Irregular articular surfaces & 12 & 5 & 2 \\
& & Joint subluxation or dislocation & 30 & 14 & 22 \\
& & Intra-articular loose body & 8 & 1 & 2 \\
& & Others & 5 & 12 & 2 \\
\hline

\multirow{10}{*}{Muscle} & \multirow{10}{*}{Muscle} & Muscle swelling & 8 & 5 & 1 \\
& & Muscle atrophy & 13 & 1 & 0 \\
& & Muscular soft tissue mass & 40 & 9 & 2 \\
& & Muscular cystic mass & 11 & 13 & 3 \\
& & Muscular hemorrhage & 7 & 0 & 1 \\
& & Muscle calcification & 3 & 0 & 0 \\
& & Muscle open injury and tear & 0 & 0 & 0 \\
& & Tendon calcification & 19 & 3 & 13 \\
& & Tendon tear or rupture & 1 & 0 & 1 \\
& & Others & 7 & 4 & 1 \\
\hline

\multirow{10}{*}{\makecell{Skin and \\subcutaneous fat}} & \multirow{10}{*}{\makecell{Skin and \\subcutaneous fat}} & Subcutaneous soft tissue mass & 141 & 47 & 28 \\
& & Subcutaneous edema & 18 & 6 & 2 \\
& & Subcutaneous effusion & 26 & 3 & 1 \\
& & \cellcolor[gray]{0.95}Subcutaneous inflammatory & \multirow{2}{*}{10} & \multirow{2}{*}{2} & \multirow{2}{*}{1} \\
& & \cellcolor[gray]{0.95}exudate & & & \\
& & Subcutaneous swelling & 60 & 10 & 15 \\
& & \cellcolor[gray]{0.95}Subcutaneous open wound & \multirow{2}{*}{15} & \multirow{2}{*}{5} & \multirow{2}{*}{2} \\
& & \cellcolor[gray]{0.95}and laceration & & & \\
& & Subcutaneous calcification & 12 & 0 & 0 \\
& & Subcutaneous fat necrosis & 2 & 0 & 0 \\
& & Abdominal wall hernia & 48 & 3 & 6 \\
& & Others & 33 & 5 & 9 \\
\hline

\multirow{4}{*}{Agenesis or ectopia} & \multirow{4}{*}{Agenesis or ectopia} & \cellcolor[gray]{0.95}Congenital developmental & \multirow{2}{*}{78} & \multirow{2}{*}{35} & \multirow{2}{*}{14} \\
& & \cellcolor[gray]{0.95}anomaly & & & \\
& & Postoperative changes & 42 & 16 & 5 \\
& & Situs inversus & 10 & 7 & 1 \\
& & Others & 2 & 0 & 0 \\
\hline

\multirow{2}{*}{\makecell{Implantation of \\artificial object}} & \multirow{2}{*}{\makecell{Implantation of \\artificial object}} & Implantation of artificial object & 179 & 91 & 53 \\
& & Others & 17 & 3 & 8 \\

\bottomrule
\end{tabular}
% }
\caption{(Continued) Detailed distribution of abnormality categories in \dataset~.}
\label{tab:detailed_distribution10}
\end{table*}

% The followings are for CVPR

% \input{sec/0_abstract}    
% \input{sec/1_introduction}
% \input{sec/2_relatedworks}
% \input{sec/3_taxonomy}
% \input{sec/4_dataset}
% \input{sec/5_method}
% \input{sec/6_experiments}
% \input{sec/7_conclusion}
% \clearpage
% {
%     \small
%     \bibliographystyle{ieeenat_fullname}
%     \bibliography{main}
% }
% \clearpage
% \input{sec/X_suppl}

\end{document}